\documentclass{ws-ijmpd}
\usepackage[super,compress]{cite}
\usepackage{url}
\usepackage{hyperref}
\usepackage{graphicx}
\usepackage[usenames]{color} 
\usepackage{amssymb}
\usepackage{amsmath}
\usepackage[normalem]{ulem}
\usepackage{xspace}
\usepackage{ulem}
\RequirePackage{epsf}
\newcommand{\be}{\begin{equation}}
\newcommand{\ee}{\end{equation}}

\newcommand{\etal}{{\it et al.}}
\newcommand{\mth}{\ensuremath{M_{\rm thres}}}
\newcommand{\bsube}{\begin{subequations}}
\newcommand{\esube}{\end{subequations}}
\newcommand{\bea}{\begin{eqnarray}}
\newcommand{\eea}{\end{eqnarray}}
\newcommand{\beaa}{\begin{eqnarray*}}
\newcommand{\eeaa}{\end{eqnarray*}}
\usepackage{float}
\begin{document}


\markboth{Authors' Names}
{ASTROPHYSICAL IMPLICATIONS OF NEUTRON STAR INSPIRAL AND COALESCENCE}

%
\catchline{}{}{}{}{}
%

\title{BASIC ASTROPHYSICS AND ASTROPHYSICAL IMPLICATIONS OF NEUTRON STAR INSPIRAL AND COALESCENCE}

\author{JOHN L. FRIEDMAN}

\address{Department of Physics, University of Wisconsin-Milwaukee\\
Milwaukee, Wisconsin 53217, US\\
friedman@uwm.edu}

\author{NIKOLAOS STERGIOULAS}

\address{Department of Physics, Aristotle University of Thessaloniki\\
  Thessaloniki, 56124, Greece\\
niksterg@auth.gr}

\maketitle

\begin{history}
\received{Day Month Year}
\revised{Day Month Year}
\end{history}

\begin{abstract}
The first inspiral of two neutron stars observed in gravitational waves was 
remarkably close, allowing the kind of simultaneous gravitational wave 
and electromagnetic observation that had not been expected for several years.  
Their merger, 
followed by a gamma-ray burst and a kilonova, was observed across the spectral 
bands of electromagnetic telescopes.  These GW and electromagnetic observations 
have led to dramatic advances in understanding short  
gamma-ray bursts; determining the origin of the heaviest elements;  and 
determining the maximum mass of neutron stars.  From the imprint of tides on 
the gravitational waveforms and from observations of X-ray binaries, 
one can extract the radius and deformability of inspiraling neutron stars.  
Together, the radius, maximum mass, and causality constrain 
the neutron-star equation of state, and future constraints can come  
from observations of post-merger oscillations.  We selectively review these results,
filling in some of the underlying physics with derivations and estimates.  
 
\end{abstract}

\keywords{Keyword1; keyword2; keyword3.}

\ccode{PACS numbers:}
\section{Introduction}

       Observing a 100-second train of gravitational waves (GWs) from inspiraling 
neutron stars\cite{Abbott2017} followed after 1.7 s seconds by a gamma-ray burst\cite{dnsgrb} 
immediately and incompletely resolved the 50-year old mystery of short gamma-ray bursts (sGRBs).  
Within 11 hours, afterglow light was seen from a host galaxy 130 Ly away, implying 
that the speed of gravitational waves agreed with the speed of light to a few seconds 
in 130 My, one part in $10^{15}$.  Over the next two years, spanning photon energies 
from $10^{-5}$ eV to $100$ MeV, telescopes across the world monitored the post-merger 
light.  These observations, together with numerical simulations of mergers, elucidate 
the nature of the afterglow and give compelling evidence that a substantial fraction 
of the universe's heaviest elements are forged by rapid neutron bombardment of nuclei 
ejected in the merger.  

They also imply that a massive post-merger neutron star briefly sustained itself 
against collapse. This in turn, leads to a new lower limit on the maximum neutron star mass 
and the most precise current estimate of its value.  The primary observational
constraints on the behavior of cold matter above nuclear density -- on the neutron-star 
equation of state -- are this maximum mass and measurements of neutron star masses and 
radii.  In the late inspiral, tides alter the waveform; by comparing the GW170817 inspiral 
waveform to a template bank of waveforms developed in years of analytic and numerical studies, 
the Ligo/VIRGO collaboration (LVC), as well as subsequent authors, measured the masses and 
radii of the two neutron stars to within about 2 km.  These and future GW measurements 
supplement electromagnetic observations of neutron stars in binary systems that have given 
precise measurements of mass and approximate measurements of radius. 

Observed neutron stars with masses above 2$M_\odot$ and the strong evidence from 
GW170817 that the maximum mass is above 2.1$M_\odot$ are close to ruling out the hypothesis 
that neutron stars are really strange quark stars.  The evidence makes it significantly less likely 
that neutron stars have quark cores and reduces the likelihood of cores with hyperons;\cite{Alford2015} 
the constraints on EOS parameters are not, however, stringent enough to rule out either 
alternative.\cite{Annala2019}
 
We selectively review the implications for physics of the inspiral and merger of two neutron stars. 
Papers on astrophysical implications of NS-NS merger typically use models and relations whose 
derivations must be tracked through the literature. In our discussion, 
we fill in some of the physics, with calculations and estimates. 
  
The plan of the paper is as follows.  In Secs.~\ref{s:merger} and \ref{s:r-k}, we discuss the merger of binary 
neutron stars and key parts of the physics underlying gamma-ray bursts,  
kilonovae, and the creation of r-process elements.  We turn in Sec.~\ref{s:radius} to the 
determination of NS radius and tidal deformability from binary inspiral and from electromagnetic 
observations; and in Sec.~\ref{s:mass} to inferring maximum NS mass from inspiral and postmerger 
observations and from X-ray binaries.  In Sec.~\ref{s:eos}, we consider the implications for the 
EOS of neutron star matter of radius and maximum-mass observations and of future observations 
of post-merger oscillations of a hypermassive star prior to collapse. Finally, in Sec.~\ref{s:hubble}, we 
give a simple derivation of Schutz's method of finding Hubble's constant from the inspiral waveform 
of a source of known redshift.  

Among the recent, more detailed expositions are a Shibata-Hotokezaka review of merger and mass ejection,\cite{sh19} 
reviews of kilonovae by Metzger\cite{Metzger2019}, of gamma-ray bursts by Berger\cite{Berger2014} and {M\'esz\'aros,\cite{Meszaros2019}, and of 
observational and theoretical constraints on the neutron-star EOS by Lattimer.\cite{Lattimer2017}

\section{Mergers of compact binaries and short gamma-ray bursts} 
\label{s:merger}

In the merger of a double neutron star (NS-NS) system, if the total mass $M$ is small 
enough to temporarily sustain itself against collapse, the system must rid itself 
of the difference between the orbital energy just prior to merger and the kinetic energy of 
a differentially rotating remnant.  

A rough estimate of the size of a  $2.7M_\odot$ remnant is obtained assuming an average density equal to nuclear density, 
$\rho_n = 2.6\times 10^{14}\rm\ g/cm^3$: 
$ \displaystyle  R \sim \left[2.7 M_\odot/(\frac43\pi\rho)\right]^{1/3}\sim 17 \ \rm km$. 
(this roughly agrees with numerical simulations, when low-density material surrounding the remnant is included). The available energy is then of order the gravitational binding energy        
\be
  E \sim \frac1{10}\frac{GM^2}{R}\sim 10^{53}\left(\frac M{2.7 M_\odot}\right)^2\frac{15\ \rm km}{R}\  \rm erg,
\ee 
an energy per baryon of order 50 MeV.  
Almost all the energy of the merger is ultimately carried away by 
gravitational waves and by thermal neutrinos. These are produced by electron and positron capture, 
$
        e^+ + n\rightarrow p+\bar\nu_e, \ p+e^-\rightarrow n + \nu_e\ 
$,
by pair annihilation, $e^+ +e^-\rightarrow \nu+\bar \nu$, and by electrons interacting with the hot plasma 
$
    e^-\rightarrow e^-+\nu+\bar\nu,
$ 
the latter processes yielding all neutrino flavors.  
The remaining energy is in the kinetic energy of ejected material, in kinetic and thermal energy 
of the remnant, in a growing magnetic field, and in light.  Because only about 1\% of the 
available energy is emitted as light, the features of electromagnetic observations are 
sensitive to initial conditions, 
but even the gross behavior of the system depends on the initial masses and on the 
neutron-star equation of state.  

Most important is the total mass $M=m_1+m_2$ of the system.  The nature and final fate 
of the merger depends on whether $M$ is larger or smaller than three critical masses: 
\vspace{-3mm}
\begin{enumerate}
\item \mth\  is the threshold mass, the mass above which the merged stars -- the merger {\rm remnant} --
promptly collapses to a black hole; it is the maximum mass that can be supported against 
collapse by pressure and differential rotation of the initial hot remnant.  
\item $M_{\rm max,rot}$ is the maximum mass of a cold, uniformly rotating neutron star. 
$M_{\rm max,rot}$ is about 20\% larger than 
\item $M_{\rm max, spherical}$, the maximum mass of a cold, nonrotating star, which we will denote by $M_{\rm max}$ for simplicity.  
\end{enumerate}

If $M> \mth$, the system will promptly collapse to a black hole.  The value of $\mth$ depends on 
the neutron star equation of state (EOS), and as we discuss below, measuring \mth\ 
gives the strongest astrophysical constraint on the EOS at several times nuclear density.  

When $M>\mth$, the star collapses in a few dynamical times, where  
\be
   t_{\rm dynamical} \sim \sqrt{\frac{R^3}{GM}}
        = 0.12 \left(\frac R{17 \ \rm km}\right)^{3/2} \left(\frac{M}{2.7M_\odot}\right)^{-1/2} \ \rm ms.
\ee
Even in this prompt collapse, some matter can remain in a disk, if the tidal torques are large 
enough.  The height of tides raised on the less massive star $m_1$ is roughly
\be
   h\sim \frac{m_2}{m_1}R_1\left(\frac{R_1}d\right)^3
\ee
with $R_1$ the radius of $m_1$ and $d$ the distance between the stars.  For approximately equal 
masses, the disk mass is negligible, but if $m_1/m_2 \lesssim 0.8$ and the EOS is stiff below 
twice nuclear density (so that the radius of $m_1$ is not too small), tidal disruption 
can leave a disk with mass $\gtrsim 0.01 M_\odot$.\cite{sh19}
 
\begin{figure}[h!]
\centerline{    \includegraphics[width = 10cm]{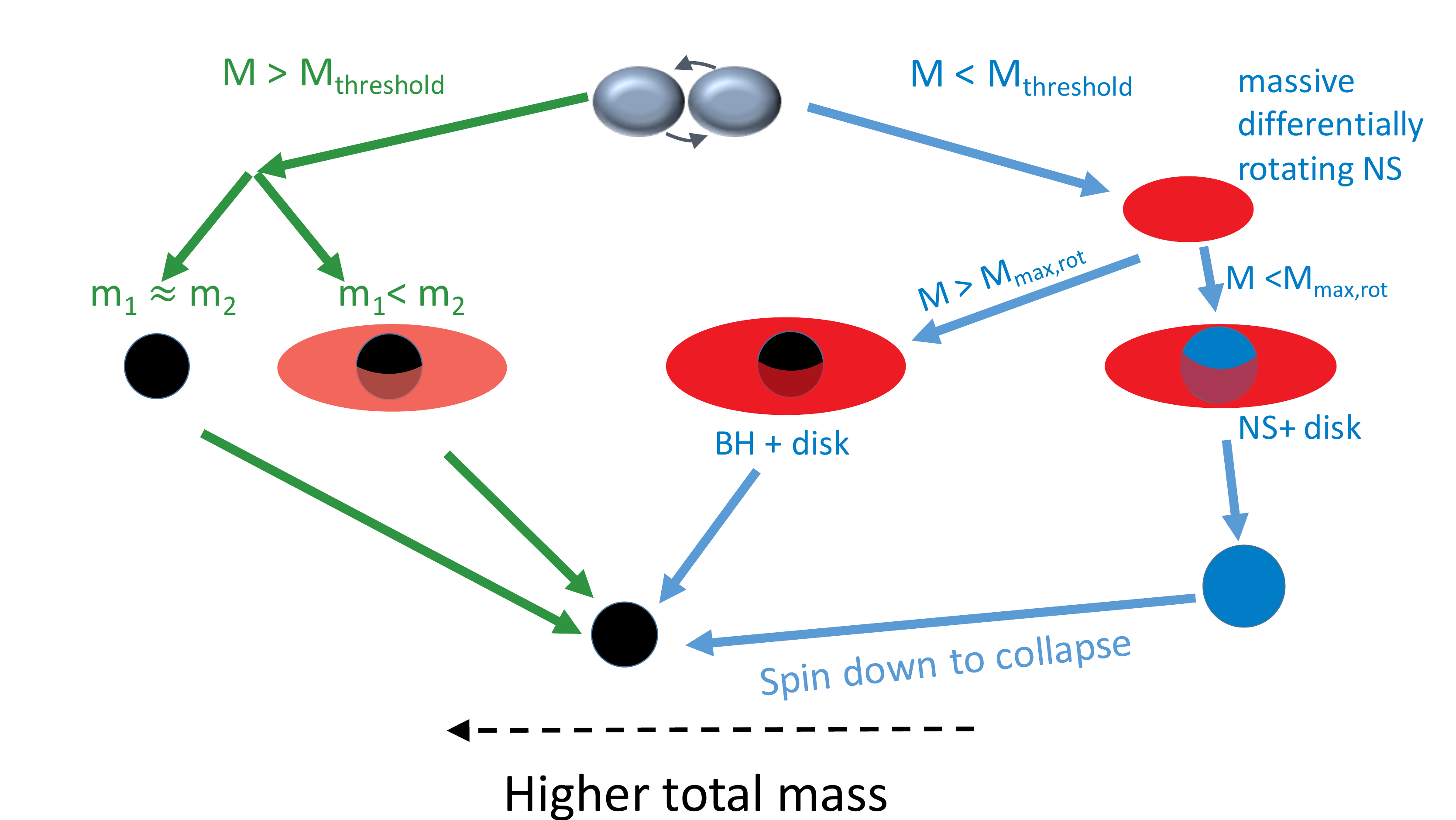}}
        \caption{The merger outcome depends on the system's total mass. $M>\mth$ leads to 
prompt collapse with a small or negligible disk.  $M< \mth$ yields a massive hot, differentially rotating 
neutron star. If $\mth>M> M_{\rm max, rot}$ this differentially rotating remnant collapses 
to a black hole as viscosity and a magnetic field enforce uniform rotation. 
Figure adapted from Shibata and Hotokezaka\cite{sh19} }
\label{fig:merger}\end{figure}

When $M<\mth$, the merger yields a hot, massive neutron star surrounded by a thick disk of material.  
The massive remnant is initially supported against collapse by differential rotation and degeneracy pressure (the pressure of the zero-temperature neutron star EOS), 
which still dominates thermal pressure in the dense core.  At the intersection of the merging stars, 
velocity fields from each star are oppositely oriented, leading to a differentially rotating remnant
and to an unstable boundary layer between fluids with different velocities.  This is the Kelvin-Helmholtz instability, responsible, for example, for ocean waves -- 
waves at the interface between air and water.   Here the length scale is much shorter than the 
radius of the star, and the growth time much faster than dynamical.  Differential rotation also leads to 
rapid magnetic field growth, in part through the magnetorotational instability.\cite{chandra61,bh94}

The ensuing gross behavior over the next several seconds is tied to a redistribution of angular momentum.  
A ring of fluid with mass $m$, radius $r$, and velocity $v$ has the same angular momentum 
$mvr$ as a more distant ring at radius $R$ and velocity $vr/R$, but the energy of the more 
distant ring is smaller by the factor $r^2/R^2$.  Dissipation -- here viscosity and 
magnetic turbulence -- therefore transports angular momentum outward, driving the rotation 
law toward uniform rotation. The smaller amount of kinetic energy available in  
uniform rotation leads to the collapse that differential rotation had temporarily halted:  
A star is called {\sl hypermassive} if its mass is in the range $\mth > M > M_{\rm max,rot}$, 
below the threshold for prompt collapse, but too large to be supported by uniform rotation.
\cite{Baumgarte2000}

Finally, even if $M< M_{\rm max,rot}$, collapse remains a short-term threat, because   
the newly generated magnetic field will spin the star down, and radiation will cool its 
outer part.  The spin-down time from a magnetic field of order $10^{15}$ G is a few minutes, and 
unless the remnant's mass is below the maximum mass $M_{\rm max}$ of a cold, 
nonrotating star, the merging stars are fated to end as a black hole surrounded 
by a disk (see, e.g., \cite{sh19} and references therein).

\subsection{Short gamma-ray bursts}
\label{s:sgrb} 

    Beginning in the mid-sixties with observations of the Vela satellites, bursts of gamma rays 
were detected and subsequently identified with galaxies at cosmological distances. There are 
two primary classes:  Long bursts, with duration generally greater than two seconds,  
occur in star-forming galaxies, and coincident supernovae are seen when the burst is 
close enough for the supernova to be detected.  They are compellingly linked to the core-collapse of a 
massive star, ending as a black hole or possibly a magnetar (a neutron star with a 
magnetic field of order $10^{15}$ G).  For short gamma-ray bursts, on the other hand, 
no associated supernova has been seen, and, when associated galaxies are seen, they often 
have only old stars, ruling out supernovae as the source.  Because the bursts last less 
than 2 s, they must emerge from a volume that is not much more than 2 light-seconds across.
The cosmological distance implies a luminosity comparable to the neutron-star binding 
energy seen in supernovae. With supernovae ruled out, the leading candidates have been 
NS-BH and NS-NS mergers\cite{paczynski86,nps91}  (see, e.g., \cite{Meszaros2019,Berger2014} for 
a history and extensive references).%

   The gamma-ray burst is emitted by a jet of relativistic particles that fly outward along the 
axis of rotation.  The restriction of the burst to a narrow beam is associated with 
relativistic beaming and to collimation of the jet by pressure of the surrounding matter 
and perhaps by a toroidal magnetic field.\cite{Granot2015,Bromberg2018}
A leading candidate for the engine that launches the jet is the Poynting flux from a magnetic 
field of strength above $10^{15}$ G that accelerates plasma near the 
axis to relativistic speeds.  Driven by the Kelvin-Helmholtz instability and the magnetorotational instability (MRI) associated with differential rotation, the magnetic field grows dramatically on a dynamical timescale. Recent general relativistic magnetohydrodyamical simulations of NS-NS mergers 
with magnetic field growth are reported by Rezzolla {\sl et al.}\cite{Rezzolla2011}, Kiuchi {\sl et al.}, \cite{Kiuchi2014} and by Ruiz  {\sl et al.} \cite{Ruiz2017}, who find the emergence of a mildly relativistic jet.  Because of its astrophysical importance, we briefly discuss the MRI and estimate its growth time, following Balbus and Hawley.\cite{bh91}. 

The post-merger star inherits a small magnetic field from its progenitors. Because the matter is 
hot and conducting, a displacement $\xi$ of the fluid deforms the magnetic field.  
The energy per unit volume of the magnetic field is $\frac{B^2}{8\pi}$, and bending the field lines to a curve with radius of curvature $R$ then gives a force per unit volume of order 
$B^2/R$. 
Because a displacement $\bm\xi \cos({\bm k\cdot\bm x})$ has radius of curvature $\lambda^2/\xi$,
the restoring force per unit volume for a displacement perpendicular to $\bm B$, with $\bm k$ along $\bm B$, is 
 $   f \sim \xi B^2/\lambda^2$. 

The way the instability works can be seen in a toy model, with the star replaced by a 
disk about a central mass and each particle in the disk moving with its Keplerian 
angular velocity $\Omega = \sqrt{GM/r^3}$.  A particle displaced outward by an amount $\xi$ 
(keeping its initial angular momentum) moves in an elliptical orbit with the same frequency $\Omega$: 
In a frame moving with the original angular velocity $\Omega$, the particle oscillates 
about its original position with oscillation frequency (epicyclic frequency) $\Omega$, corresponding 
to a restoring force per unit volume $\rho\Omega^2\xi$.     

Now consider the same model with a magnetic field perpendicular to the disk.  When a 
fluid element is displaced outward, its angular velocity decreases; in the rotating frame 
it moves backward, and it drags the field lines backward.  The deformed field pulls the fluid 
element forward, countering the decrease in its angular velocity.  If the resulting 
angular velocity at $r+\xi$ is larger than $\Omega(r+\xi)$ of the unperturbed disk, 
the fluid element's acceleration will be larger than the gravitational force on it 
and it will accelerate outward. The criterion for instability is then that the 
magnetic force per unit volume is larger than the restoring force per unit volume or, roughly,%
   $\displaystyle\frac{B^2}{\lambda^2} \gtrsim \rho\Omega^2$.
Modes with wavelength of order $R$ 
are unstable on a dynamical time when $B^2\sim \rho\Omega^2 R^2$.  Thus, energy 
associated with differential rotation can flow into a magnetic field on a 
dynamical timescale.  The instability criterion for a differentially rotating star  
has this form, as long as $d\Omega/dr \sim -\Omega/R$, and it is satisfied in the 
the outer part of the massive neutron star.\cite{Hanauske2017}  The inner part 
can retain two density maxima and a quadrupole velocity field from the merger until collapse; 
the MRI is not confined to the outer part of the star, but one cannot use the criterion for 
a differentially rotating star.  

Short wavelength magnetic turbulence and viscosity convert energy associated with differential 
rotation in the outer part of the remnant to heat and kinetic energy.  
The available energy is a fraction of the difference $\Delta E$ between the initial 
differential rotation and a final uniform rotation, and the change in angular velocity is 
comparable to the angular velocity itself.  Estimating the moment of inertia as 
$I\sim\frac25 MR^2$, with $M=2.6 M_\odot$ and $R=14$ km,%
\footnote{Despite the fact that any spherical Newtonian star has moment of inertia smaller than
$I=\frac25 MR^2$ (the uniform-density case), the enhanced strength of relativistic gravity at 
high compactness increases the value of $I$ for the same $M$ and $R$. 
Following Bejger \& Haensel\cite{bh02}, Lattimer \&\ Schutz\cite{ls05} find 
$I\approx 0.237 MR^2\left[1+4.2 (M/M_\odot)/(R/1\rm km) + 90 (M/M_\odot)^4/(R/1\rm km)^4\right]$, 
to about 3\%. }
we have
\be
   \Delta E \sim \frac12 I\Omega^2 \sim 10^{53} \left(\frac I{4\times 10^{45}\ \rm g\, cm^2}\right)\left(\frac{\Omega}{7000\rm\ s^{-1}}\right)^2\ \rm erg.
\label{e:DeltaE}\ee   
With $\sim 10^{57}$ baryons per $M_\odot$, and $\lesssim 0.05 M_\odot$ in the ejecta 
this energy is over $1000$ MeV/(ejecta baryon). 

Unless $M<M_{\rm max,rot}$, collapse leaves a large part of the rotational energy in the black hole.  
Most of the remainder is emitted as neutrinos. Part, less than $10^{51}$ erg, powers the ejection of a thick torus extending from the equatorial plane and moving with roughly the escape velocity, about $0.1$c, 
or a kinetic energy of about 5 MeV per baryon.  The part that is ordinarily observed, the gamma-ray 
burst itself, has a typical energy less than 2$\times 10^{50}$ erg.\cite{fong15,Berger2014}%
\footnote{This energy is computed after estimating the solid angle of the beam.  Because the beam 
angle is often not known, an {\sl isotropic equivalent energy} $E_{\rm iso}$, the larger 
energy of an isotropic emitter with the same observed flux, is common in the literature: 
Emitted energy $= E_{\rm iso}\times$ (solid angle of the beam)/$4\pi$. For SGRBs, the estimated 
average relation is\cite{Berger2014} $E \sim 0.015 E_{\rm iso}$.} 
Remarkably, however, the gamma-rays 
are emitted by highly relativistic matter:  As we review in the next paragraphs, there is 
compelling evidence that the gamma-rays are emitted by a jet whose matter has a Lorentz factor 
of order 20-100 at the time of observation.  If the emitting particles were accelerated baryons, 
they would need an energy per baryon $m_n\Gamma > 20,000$ MeV.  For accelerated electrons or 
$e^+e^-$ pairs, the energy per particle $m_e\Gamma$ instead exceeds a more manageable 
10 MeV.  

  A lower limit on $\Gamma$ is set by the observation of gamma-ray energies 
above the threshold, $m_ec^2 = 511$ keV, for pair creation from $\gamma+\gamma \rightarrow e^++e^-$. 
We will soon see that, if this were the energy in the rest frame of the matter, 
the interaction of the photons with a high density of pairs would make the emitting region opaque. 
But the prompt emission of the gamma rays (and the fact that the spectrum is not a blackbody) 
means that the matter is nearly transparent, that their mean free path $\ell$ is of order 
the radius $R$ of the emitting region. 

We can estimate $\Gamma$ as follows.\cite{gfr83} For a total energy 
${\cal E}(m_ec^2)$ of light whose energy per photon is above $m_e c^2 = 511$ keV 
in the rest frame of the radiating matter, 
a fraction of order unity of the photons are converted to pairs, giving a total 
number of pairs $N\sim {\cal E}/m_e c^2$ and a number density
$
   n = \frac{\cal E}{m_e c^2 R^3},
$
with $R$ the radius of the emitting region.  

 The scattering cross section of gamma-rays from electrons (and positrons) is the 
classical Thomson cross section 
$
  \sigma_T = \frac{8\pi}3r_e^2 = 0.67\times 10^{-24}\ \rm cm^2,  
$
where $r_e = e^2/(m_ec^2)$ is the classical electron radius.  The mean free path, 
$ \ell = \frac1{n\sigma_T}$,  is then
$\displaystyle  \ell \sim \frac {m_e c^2 R^3}{{\cal E}\sigma_T}$,
and the condition $\ell\gtrsim R$ is 
\be
    \frac {m_e c^2 R^2}{{\cal E}\sigma_T} \gtrsim 1.  
\ee
If the matter is not highly relativistic, this is wildly inconsistent with observation: 
The burst is emitted in about a second from matter expanding outward with $v\approx c$, 
implying $R\sim 3\times 10^{10}$ cm.   
The total burst energy is 
${\cal E}_{\rm observed}\gtrsim 10^{48}$ erg. If the matter is not highly relativistic, this observed energy is comparable to the energy in the rest frame of the matter, 
and we have   
\be
  \frac {m_e c^2 R^2}{{\cal E}\sigma_T} \approx 10^{-9} \left(\frac R{3\times 10^5\rm km}\right)^2
                                          \left(\frac {\cal E}{10^{48}}\ \rm erg\right)^{-1}.   
\label{e:ell}\ee
That is, if $\Gamma\sim 1$, the mean free path is less than $R$ by a factor of at least $10^9$. 

If the matter is highly relativistic, however, the observed energy is much larger than the energy 
in the rest frame of the matter.  The most obvious reason is relativistic beaming, but there is
a second effect that is more important:  Because photons are blue-shifted by the factor 
$\sqrt{(1+v/c)/(1-v/c)}\approx 2\Gamma$ for $v\approx c$, the rest-frame energy  
peaks at a photon energy $E_p = E_{p,\rm observed}/2\Gamma$, leaving far fewer photons 
with energy high enough for pair creation.  Beyond the peak energy, the observed energy
decreases, either as a power law or an exponential (see, e.g., a review by Nakar\cite{Nakar2007}).
We obtain a conservative 
lower limit (an underestimate) with an exponential cutoff: With ${\cal E}(E)$ the 
energy greater than $E$, the cutoff has the form%
\footnote{As reviewed in Nakar\cite{Nakar2007} the range of SGRBs are fit to models of the 
form ${\cal E}(E) = A E^{\alpha+1} e^{-E/E_0}$, with $\alpha$ in the range $-2$ to 
$0.5$ (other models use broken power laws).  We chose a representative value 
$\alpha=-1$, with $A=1$ to give a total energy of order $E_0$. Observed peak energies 
range from 20 to 1000 keV.\cite{ggc04}}      
$\displaystyle {\cal E}_{\rm observed}(E) = {\cal E}_{\rm observed}(E_p) \exp(-E/E_{p})$,
with $E_{p,\rm observed}= 500$ keV a typical value.  The energy distribution in the 
matter rest frame is then given by
\be
   {\cal E}(E) = \frac{{\cal E}_{\rm observed}(E_p)}{2\Gamma} \exp(-2\Gamma E/E_p),
\ee 
and Eq.~\eqref{e:ell} for ${\cal E} = {\cal E}(m_ec^2) $ gives $\ell> R$ 
for $(2\Gamma)^{-1}e^{-2\Gamma} < 10^{-9}$, or $\Gamma > 9$. SGRBs have varying cutoffs and peak energies, and estimated values of $\Gamma$ are larger than our deliberate underestimate, 
ranging from 20 to 100.  

Without the effect of the energy cutoff, relativistic beaming and blue-shifting  
alone enhance the total observed energy of the burst by a factor of $\Gamma^3$. 
The factor of $10^{9}$ needed for $\ell\sim R$ would then require $\Gamma\gtrsim 10^3$.   

As mentioned above, magnetic transfer of rotational energy in merger simulations 
has produced only mildly relativistic outflows. A competing central 
engine, neutrino pair annilation $\nu+\bar\nu\rightarrow e^+ + e^-$ followed by $e^\pm$ 
annihilation to produce gamma rays, also appears to provide too little energy.%
\cite{mr92,Ruffert1997,DiMatteo2002,Birkl2007,Dessart2009,Perego2017a,Siegel2017}
Following collapse, however, 
the spinning black hole will retain a magnetic field that can again communicate  
its rotational energy to the surrounding plasma (the Blandford-Znajek mechanism\cite{BZ77}) 
and that may provide the engine that turns the outflow into a highly relativistic jet.%
\cite{McKinney2006,Rezzolla2011,Kiuchi2014,Ruiz2016}

\vskip0.3cm

\noindent{\sl The Burst GRB170817A} \\     

The gamma-ray burst associated with GW170817 was unusually weak, with 
luminosity less than $4\times 10^{47}$ erg s$^{-1}$, orders of magnitude below 
that of a typical sGRB\cite{fong15} with known redshift.   A debate over whether 
the jet had made it past the surrounding matter or was choked was finally resolved by 
direct observation\cite{Mooley2018,Ghirlanda2019} of ``superluminal motion,'' a source that moved a 
measured distance $d$ perpendicular to the line of sight in a measured time
less than  $d/c$.  Recall that, for a source moving at $v\approx c$ at 
an angle $\theta$ to the line of sight, light emitted after a time 
$t$ travels a distance shorter by $vt\cos\theta$ than light emitted at $t=0$.
The difference in arrival time is then $\displaystyle t-vt\cos\theta/c$,   
and the apparent velocity perpendicular to the line of sight is
$\displaystyle  v_{\rm app} = \frac{v\sin\theta}{1-v\cos\theta/c}$.
This is large for small $\theta$ and has maximum value $v_{\rm app}=\Gamma c$ 
at $\theta=1/\Gamma$.  The observed apparent velocity of the late-time jet 
implies $\Gamma\approx 4$.  This is consistent with an initial $\Gamma\gtrsim 10$, 
and does not rule out a larger value.    

Besides the much weaker gamma-ray luminosity are other striking observational differences. 
An initial 0.5 s burst was followed by an unusual tail of lower-energy thermal gamma-rays lasting 
about 2 s.\cite{fermi17}  The longer-wavelength afterglow was also idiosyncratic: 
In earlier events, as interactions with surrounding matter slow the highly relativistic initial jet, its average photon energy quickly decreases, moving from gamma-rays to X-rays. Measured by time in the 
rest-frame of a jet, X-rays in typical sGRBs are seen within two minutes of the initial burst, 
but it was 9 days after the GW170817 merger before the first evidence of X-rays,\cite{Fong2017} 
and radio emission was was similarly delayed. 

Scenarios that reproduce the afterglow observations, from X-ray to radio, all involve a burst 
that is viewed from an angle of at least 15$^\circ$ from the jet axis (see \cite{Nakar2020} for a 
summary and analysis of the estimates).  Because relativistic 
beaming concentrates most of the energy inside a cone of angle $1/\Gamma$, the observed 
luminosity outside the jet cone is sharply reduced.      
A key remaining question is whether the larger viewing angle is the primary reason 
for the differences between this event and typical sGRBs or whether the event itself 
is atypical.   

In both cases, however, the description of the observed gamma-ray burst and subsequent 
rise and fall of the afterglow can be described by a standard scenario in which a relativistic 
jet, because its speed is far above the sound speed of the ejecta, generates shock waves that 
dissipate some of its energy to a surrounding sheath, called a cocoon (the 
term is from the early Blandford-Rees model of galactic jets\cite{br74,Scheuer1974}). 
Because the viewing angle is larger than the opening angle of the jet, 
the observed burst and the early afterglow come from the mildly relativistic cocoon. 
As the jet loses energy and the beaming angle $1/\Gamma$ increases, light from the jet's core 
ultimately becomes visible and dominates the observed luminosity. The resulting appearance of 
the gamma-ray burst and its afterglow are described as a jet+cocoon or 
as a stuctured jet, a term that avoids an assumed underlying mechanism.
The ambiguity in whether 170817 is a typical sGRB comes from the fact that, the more distant 
sGRBs are seen on-axis: One sees the central jet emission, but not the early 
(presumably cocoon) contribution to the afterglow; 
and in 170817, the situation is opposite: one sees early cocoon emission, and light from a 
central jet is seen only at times too late for it to be visible in the more distant sGRBs. 

In arguing for a scenario in which the burst itself is atypically weak,\cite{Hallinan2017,Gottlieb2018,Mooley2018,np18,Bromberg2018} 
the authors typically find that a lower limit on the jet's Lorentz factor $\Gamma$ 
of order 10 is consistent with the remaining observations.  They do not in general 
exclude a larger $\Gamma$ for the unobserved initial jet.     
Authors arguing that GW170817 is typical\cite{Fong2019,Wu2019,Salafia2019,Lazzati2018} 
emphasize the agreement 
of the late afterglow, when light from the widening jet core is dominant, with the 
late-time appearance of on-axis sGRBs: In particular, as the jet slows and widens, 
the luminosity of sGRBs decreases roughly as $1/t^2$ to $1/t^{2.5}$  
), and the late 170817 decline (although not visible until much later than that of on-axis sGRBs) 
fits that power-law.  In general, they do not require a highly relativistic jet, but 
Wu and MacFayden\cite{Wu2019}, matching observations to a parametrized model, claim 
an initial Lorentz factor $\Gamma\sim 150$, well within the range 
of the on-axis sGRBs.      

If GW170817 is a typical sGRB, one might hope to find among the archived sGRB data evidence 
of overlooked events with a similar gamma-ray behavior, a sharp spike, followed by a delayed, 
lower-frequency thermal tail.  In the last year, a search by von Kienlin{\sl et al.}\cite{Kienlin2019} found 12 candidates, including one noticed a year earlier with a measured redshift.\cite{Burns2018,Troja2018}  In this last case, GRB 150101B, the observed burst energy is $10^3$ larger than 
170817, but still at the low end of sGRBs.  An analysis by Nakar \& Piran\cite{np18}, however, 
finds that at most two of the candidates are consistent with the cocoon shock-breakout model, 
suggesting that the structure of the 170817 jet is atypical.

\section{r-process nucleosyntheis and kilonovae}
\label{s:r-k}

The major media story following GW170817 was a ``clash of neutron stars forges gold.''\cite{wsj17}
We briefly review the formation of the heaviest elements from rapid neutron bombardment and 
the evidence that merger ejecta from NS-NS and perhaps NS-BH mergers 
are the sites of this nucleosynthesis.

\subsection{r-process nucleosynthesis}
\label{s:r-rprocess}

Nucleosynthesis in stellar cores proceeds in thermodynamic equilibrium, forging elements with 
successively greater binding energy per nucleon.  As Fig.~\ref{fig:nuclear_binding} illustrates, 
the binding energy per nucleon trends upward from helium through the iron group 
(baryon number $A\sim 50-60$), because larger nuclei have a smaller fraction of surface nucleons, 
of nucleons with unsaturated bonds. But the binding energy per nucleon cannot exceed its value for 
full saturation, while the Coulomb repulsion per nucleon grows as $Z^2/A$. 
\begin{figure}[h!]
\centerline{    \includegraphics[width = 8cm]{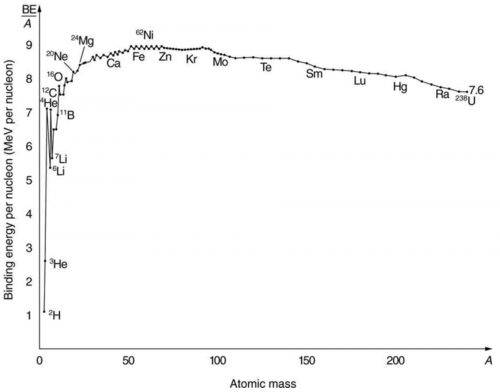}}
        \caption{Binding energy per nucleon peaks at the iron group. From OpenStax under Creative Commons} 
\label{fig:nuclear_binding}\end{figure}
After the iron group, 
the increasing Coulomb energy reverses the trend, gradually reducing the binding energy per nucleon as 
$Z$ increases.  

The result is that only elements up to the iron group can form in thermodynamic equilibrium. 
Stable nuclei, however, extend up to lead (Z=82, A=208). They lie in 
a valley of stability, limited on the high-proton side by Coulomb repulsion and on the 
high-neutron side by the increased neutron Fermi energy, and these heavier elements 
are almost entirely formed by neutron bombardment.\cite{b2fh57}  What nuclei are reached depends  
on whether the time between neutron captures is longer or shorter than the time for unstable 
nuclei to decay back to stability, ordinarily by $\beta$-decay.
Slow bombardment, the {\sl s}-process, 
proceeds through stable nuclei until an unstable nucleus is reached that then decays back to 
stability before the next neutron capture.  As $A$ increases, however, a smaller fraction of 
nuclei are stable.  Not all stable nuclei can be reached by a path through the valley 
of stability, and, more often, along the available paths, capture cross sections limit the 
abundance of s-process elements.   

With rapid bombardment, the {\sl r}-process, highly unstable neutron-rich nuclei are built 
before they have time to decay.  After the 
bombardment stops, a series of decays take the nuclei back to stability. The {\sl s}-process and 
{\sl r}-process nuclides comprise overlapping sets.  In Fig.~\ref{fig:r_s_process}, the 
solid zig-zag line shows a segment of the {\sl s}-process through stable nuclides. 
\begin{figure}[h!]
\centerline{    \includegraphics[width = 8cm]{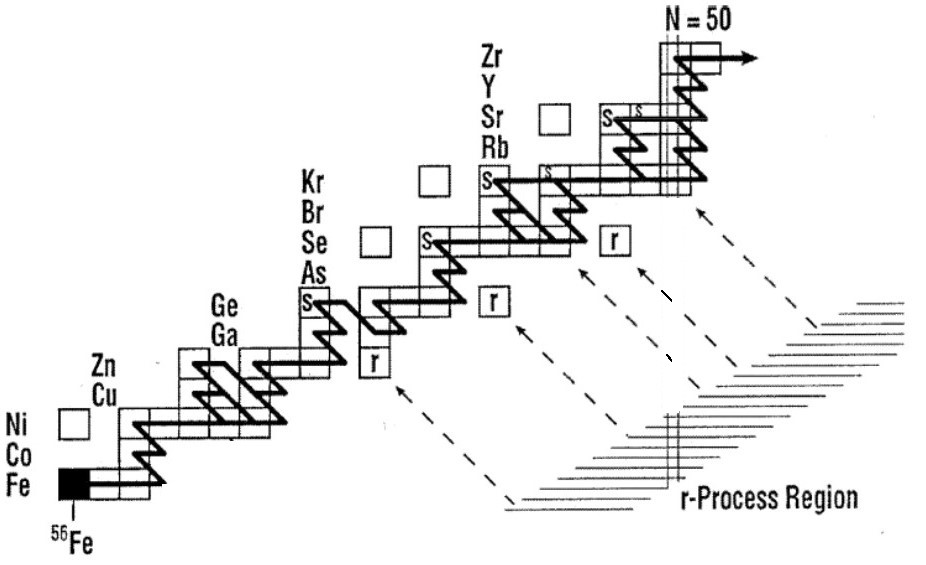}}
        \caption{Stable elements are represented by boxes, neutron number increasing to the right, 
proton number increasing vertically.  Heavy solid lines show s-process paths. 
Arrows show beta-decay to stability of initial $r$-process nuclides. Adapted from Kaeppler {\sl et al.}\cite{Kaeppeler2011}}
\label{fig:r_s_process}\end{figure}
The hashed area (labeled {\sl r}-process band) shows the unstable nuclides formed by the initial rapid 
bombardment, and the dotted lines with arrows pointing back to stability show the 
result of $\beta$-decays back to stability. 

That mergers of NS-BH and NS-NS binaries might be a primary site for $r$-process nucleosynthesis was 
first suggested by Lattimer and Schramm\cite{ls74} and by Symbalisty and Schramm\cite{symbalisty82}, respectively.  Although the astrophysics community long favored supernovae, evidence for mergers and against supernovae gradually accumulated (see Lattimer\cite{Lattimer2019} for a history of the 
debate).  The strongest observational evidence is an 
uneven distribution of $r$-process nuclides that implies they are synthesized in rare events: 
For example, of dwarf galaxies large enough to have hosted thousands of supernovae (with of order $10^6$ stars), only about 10\% have $r$-process elements, 
a percentage consistent with SGRB rates.\cite{Ji2016} On the theoretical side, in 
increasingly sophisticated supernova and merger simulations, only the mergers yield significant 
amounts of heavy $r$-process matter.     

The measured abundance of {\sl r}-process nuclides peaks at {\sl magic numbers} that correspond 
to nuclei with filled shells of neutrons, reached during rapid bombardment  
{\sl before} the nuclei decay to stability.  
The three main peaks are at $A \approx$ 80, 130, 195.
Because the abundance distribution reflects the binding energy per nucleon of the 
pre-decay nuclides, it is robust, emerging from a variety of simulations,\cite{Lippuner2017} 
when there is a wide enough range of neutron richness in the reacting matter. 
Some ejected matter must be highly neutron rich to reach the second 
and third peaks: To form elements from lanthanides (rare earth elements) through uranium,
a ratio of protons to the total number of baryons%
\footnote{This ratio is conventionally measured by the electron fraction 
 $Y_e := $ (number density of electrons)/(number density of baryons).}
less than 0.25 is
needed.\cite{korobkin_etal12,wanajo_etal14,sh19}
To reproduce the observed abundance of lighter {\sl r}-process nuclides, the ejecta must also 
include a less neutron rich component, with the ratio proton/baryon $\gtrsim 0.25$, and that is seen in simulations of mergers 
with $M<\mth$, mergers in which a massive neutron star briefly supports itself against 
collapse.\cite{wanajo_etal14,Sekiguchi2016,Goriely2015,Radice2016,Metzger2019}   

In merger simulations, the first ejecta are neutron rich.  If a high-mass neutron star is formed, 
the $e^\pm$ production and subsequent weak interactions that lead to loss of most of the merger 
energy in neutrinos also deplete the number of neutrons: The higher density of neutrons means 
a higher rate of $n\rightarrow p$ than $p\rightarrow n$ conversions, from positron and  
neutrino capture.  An additional and larger part of ejecta emerges as viscosity and a rapidly growing magnetic field convert differential rotation to uniform rotation, transporting angular momentum outward.
Indirectly powering this second part of the ejection is the energy difference $ \Delta E \sim 10^{53}$ erg of Eq.~(\ref{e:DeltaE}) in the massive neutron star and the smaller energy of differential rotation in the 
surrounding disk: Heated by what liberated energy is not radiated as neutrinos, the expanding disk 
adds to the ejecta.

\begin{figure}[h!]
\centerline{    \includegraphics[width = 10cm]{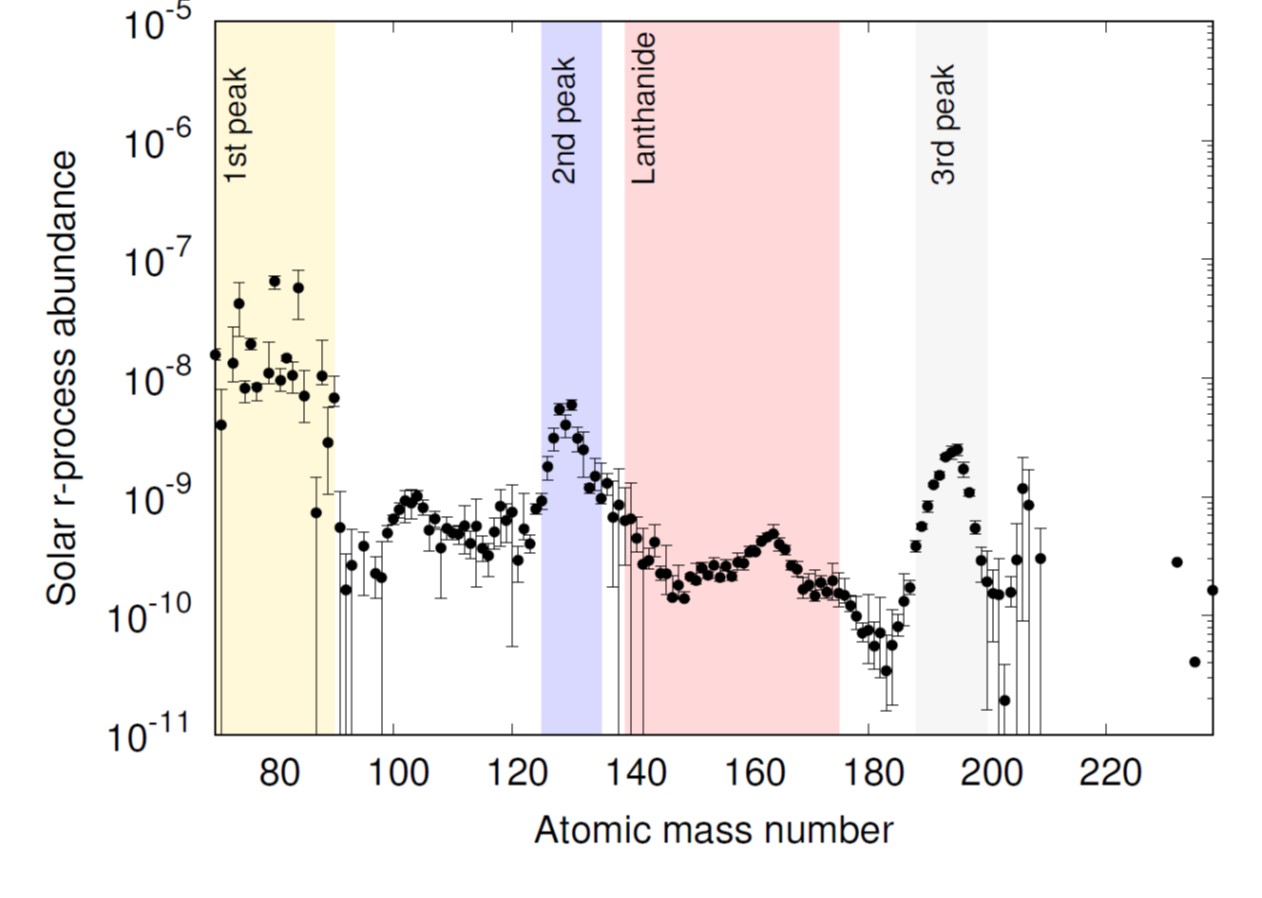}}
        \caption{Abundance of {\sl r}-process elements shows peaks associated with neutron magic numbers (filled neutron shells) 
in bombarded nuclei prior to their decay to stability. From Hotokezaka {\sl et al.}\cite{Hotokezaka2018b}}
\label{fig:abundance}\end{figure}

The LIGO-Virgo collaboration has seen one unambiguous NS-NS inspiral and coalescence (GW170817) 
and one additional event (GW190425),\cite{lvc2020} whose total mass 
is $3.4^{+0.3}_{-0.1}M_\odot$.\cite{lvc2020}, 
with a possibility that one of the component stars has a mass as large as $2.5 M_\odot$.  
If both stars are neutron stars, this second event slightly elevates the estimated rate of NS-NS 
mergers, but it may belong to the class of prompt-collapse mergers, a class unlikely to contribute 
significantly to the {\sl r}-process element abundance.  We begin with an estimate that assumes only 
one relevant event has been observed. We then suppose GW190425 is the result of a BH-NS merger and 
estimate the contribution to $r$-process matter from events of this kind.    

If, based on one event in a year of observation, we assume that a coalescence  
occurs within the observed distance at least every 5 years, we can estimate as follows an event rate 
that lies within the large error bars of the several recent estimates. The electromagnetic counterparts of GW170817 were identified with the galaxy NCG4993, whose 40 Mpc distance 
corresponds to a volume 
\be
  V = \frac43\pi(40\ \rm Mpc)^3 = 2.7\times 10^5\rm Mpc^3.
\ee
The merger rate is commonly written in Gpc$^{-3}$yr$^{-1}$ or in events per Milky-Way equivalent galaxy 
(MWEG) per year:  
\begin{align}
  \mbox{NS-NS merger rate} 
                &\sim \frac1{(5\ \rm yr)(2.7\times 10^{-4}\rm Gpc^3)} \sim 700\ \rm Gpc^{-3}yr^{-1}\\
                &\sim 0.7\times 10^{-4}\rm MWEG^{-1}yr^{-1},
\label{e:rate}\end{align}
where we have used a volume of about 100 Mpc$^3$ per MWEG.  The LIGO-Virgo rate estimate (excluding the prompt-collapse class)\cite{lvc2020} is the range $760^{+1740}_{-650}\rm Gpc^{-3}yr^{-1}$. 
(Fig. 5 of Eldridge {\sl et. al.}\cite{est19} gives a compendium of recent rate estimates 
for short gamma-ray bursts, which generally fall within this LIGO/Virgo range for NS-NS mergers). 

The observed local abundance by mass of {\sl r}-process matter\cite{Siegel2019} 
is about $10^{-7}$.
A MWEG has total stellar mass $10^{11} M_\odot$, giving a mass of $\sim 10^4 M_\odot$ of {\sl r}-process matter. 
From the properties of the electromagnetic counterpart, the ejecta 
mass for GW170817  has been estimated  to be in the range of 0.03 to 0.06~$M_\odot$\cite{Cowperthwaite2017,Kasen2017,Nicholl2017,Chornock2017,Drout2017,Smartt2017,Kasliwal2017,Kilpatrick2017,Perego2017b,Tanvir2017,Tanaka2017},
(at the high end of what is expected from numerical simulations). 
The {\sl r}-process mass per MWEG in the lifetime of the universe is then of order 
\be
   (10^{-4} \mbox{mergers/yr/MWEG})(5\times 10^{-2} M_\odot/\mbox{merger})(10^{10}\rm yr) \sim 5\times 10^4 M_\odot,
\ee
with uncertainties in the merger rate and in the {\sl r}-process production per merger allowing a range from $5\times 10^3 M_\odot$ to $10^5 M_\odot$.  If we use $0.01 M_\odot$ as a lower bound   
on the {\sl r}-process matter per merger, we have an upper limit on the  
NS-NS merger rate:  Adopting $2\times 10^4 M_\odot$ as an upper limit on the amount 
of {\sl r}-process matter in a MWEG immediately gives 
\be
\mbox{NS-NS merger rate} < 2\times 10^{-4} \rm yr^{-1}\ MWEG^{-1} 
        \approx 2000\ Gpc^{-3}yr^{-1},
\ee
below the upper edge of the LIGO/Virgo range and of the estimates 
summarized in Eldridge {\sl et al.}\cite{est19}.   

Suppose now that the event GW190425 is a BH-NS merger.  In recent simulations of a merger 
of a 2 $M_\odot$ black hole with neutron stars of radii from 11.6 to 13 km, Kyutoku {\sl et al.}\cite{Kyutoku2020} 
find tidal disruption leads to a disk of mass 0.04 $M_\odot$ to 0.1 $M_\odot$, of which 15\%-30\% is 
ejected, giving an $r$-process mass $\lesssim 0.03 M_\odot$ (but see \cite{Barbieri2020} for a larger 
estimate based on ealier simulations).  Despite a distance to the 
event four times that to  GW170817, the LIGO/Virgo paper estimates a comparable GW190425-like merger 
rate, $460_{-390}^{+1050}$ Gpc$^{-3}$yr$^{-1}$, and our previous steps then give a comparably large 
BH-NS contribution to {\sl r}-process matter: $2\times 10^4 M_\odot\ \rm MWEG^{-1}$.   
The uncertainty in ejecta mass, the very large uncertainty in event rate, the likelihood that 
this was not a NS-BH event, and the fact that one must add the unknown rate from NS-BH binaries 
with larger BH masses, all allow BH-NS events to be the dominant or a negligible contributor 
to $r$-process matter.

\subsection{Kilonovae}  
\label{s:kilonovae}

Coined by Metzger {\sl et al.} \cite{metzger10}, the term {\sl kilonova} 
refers to the peak brightness of the radioactively powered ejecta after a NS-NS or BH-NS 
merger:  The peak brightness is about 1000 times that of of a classical nova 
(the explosive fusion of hydrogen to helium at the surface of an accreting white dwarf).  
The brightness and timescale are key signatures of the formation of {\sl r}-process material  
whose radioactive decay powers the kilonova, and we give brief estimates here.  
Li and Paczynski\cite{lp98} first suggested an observable optical glow from merger ejecta 
in 1998; see Metzger\cite{Metzger2019} for a review and references. 

For a typical merger -- and, apparently, for GW170817 -- the mass is in the range 
$M_{\rm max,rot} < M < \mth$, meaning that a hypermassive neutron star forms, is supported 
by differential rotation for tens or hundreds of milliseconds 
(shorter for higher mass and/or a softer EOS),\cite{kyoto17} and collapses to a black hole as 
viscosity and magnetic field windup enforce uniform rotation.  The first, {\sl dynamical}, 
ejecta emerge immediately, in less than about 10 ms, powered by tidal disruption, by 
shocks that convert some of the collision's enormous kinetic energy to heat;\,\cite{sh19} matter at the 
interface between the merging stars is squeezed out by the collision and then swept out by 
the rotating double core.\cite{bauswein_etal13}  

The larger part of the ejecta, however, emerges later. 
As we will see, the time $t_p$ to peak brightness depends on whether or not the merger produces the 
heavier {\sl r}-process elements, in particular the lanthanides.  The time $t_p$ is much longer 
than the timescale of seconds for the larger part of the ejected matter, 
for neutrinos to carry away most of the merger energy, and for a gamma-ray burst to be launched. 
$t_p$ is long because the ejected matter is initially dense enough to be opaque -- that is, the photon mean free path is short.  It depends on whether or not lanthanides are present because 
the interaction cross section for resonant scattering or absorption grows with the number of 
available electron transitions: The lanthanides have a partly filled $f$-shell, and that leads to a large number of  available transitions. 
  
   To estimate $t_p$, note first that peak brightness occurs when photons finally diffuse 
through the ejecta with a speed $v_{\rm diffusion}$ equal to the ejecta's expansion velocity,
$v_{\rm ej}$.  Because the ejecta move at close to the escape velocity from the massive neutron star,   
$v_{\rm ej} \sim 0.1 c$.  We will find $v_{\rm diffusion}$ in terms of $t_p$ and solve 
$v_{\rm ej} =  v_{\rm diffusion}$ 
for $t_p$.  
We use
\begin{align} 
  v_{\rm ej}& = \mbox{ejecta velocity} \qquad\qquad\qquad t_p=\mbox{time from merger to peak brightness}
\nonumber\\
  \ell &= \mbox{photon mean free path}  \quad\quad\  n= \mbox{number density of ions}\nonumber \\ 
\sigma &= \mbox{interaction cross section}\qquad M_{\rm ej} = \mbox{total ejecta mass} \nonumber\\
m_S & = \mbox{average mass of ejecta ions}
\nonumber\end{align}
To find the diffusion speed, note that the time for a photon to diffuse through a radius $R$ 
is is the random walk time:  If $N=$ number of collisions, we have $R=\sqrt{N} \ell$, 
$ N = R^2/\ell^2$, implying $t_{\rm diffusion} = N\ell/ c = R^2/(\ell c)$ and 
\be
   v_{\rm diffusion} \sim \frac R{t_{\rm diffusion}} \sim \frac\ell R c.
\label{e:tp1}\ee  
The radius at peak brightness is $R=v_{\rm ej} t_p$.  To find $\ell$ in 
terms of $t_p$, write $\ell = 1/n\sigma$, where the number density is 
$n = \rho/{m_S}$, with $\rho = M_{\rm ej}/\left(\frac43\pi R^3\right)$. Then 
\be \Longrightarrow\ \frac1\ell = n\sigma = \frac{\sigma M_{\rm ej}}{m_S\left(\frac43\pi R^3\right)}.
 \label{e:tp2}\ee 
Using Eqs.~\eqref{e:tp1} and \eqref{e:tp2}, and replacing $R$ by $v_{\rm ej}t_p$, we have  
\be
  t_p  = \frac3{4\pi} \frac{\sigma M_{\rm ej}}{m_S c\, v_{\rm ej}t_p } \quad
 \Longrightarrow\ \quad  t_p = \sqrt{\frac3{4\pi} \frac{\sigma M_{\rm ej}}{m_S c\,v_{\rm ej}}}.
\label{e:tp3}\ee

  The time $t_p$ then depends on the {\rm opacity}, $\kappa = \sigma/m_S$, of order 
0.1 $\rm cm^2/g$ for ejecta with only light elements.\cite{Tanaka2019} 
For resonant interactions, the cross section $\sigma$ is enhanced by the number 
of transitions, roughly proportional to $C^2$, where the complexity $C$ is the number of ways to assign valence electrons to states 
within shells.\cite{Kasen2013} Because the lanthanides have a partially filled f-shell, 
with $2(2\ell+1) = 14$, lanthanide production from within the f-shell alone 
increases $\sigma$ by a factor of more than 100, and the total number of relevant transitions 
is an order of magnitude larger.  The lanthanides have $A\sim 150$, 
giving $\kappa = \sigma/m_S \gtrsim 10\ \rm cm^2/g$.   

Eq.~\eqref{e:tp3} then gives
\be
t_p = 7\times 10^5 \sqrt{\frac{\kappa}{10\rm\ cm^2/g}\ \frac{ M_{\rm ej}}{0.01M_\odot}\ \frac{0.1c}{v_{\rm ej}}}\quad \rm s.
\ee
Thus light escapes only after several days, when the spatial extent of the ejecta is of
order $v_{\rm ej} t_p \sim 10^{10}$ km, vastly larger than the initial post-merger configuration, 
a 20-30 km torus.      
This long time is a signature of the presence of heavy {\sl r}-process elements:
\footnote{The time is long only in the context of neutron star mergers. It is short 
compared to the timescale of supernova light curves, associated with much larger ejected envelopes 
that have little or no heavy $r$-process matter.}
Without the heavier elements, elements with valence d-shell electrons (e.g., the iron group) 
dominate the opacity, and it is smaller by a factor of 10 or more.\cite{th13,Kasen2013} 
A second key feature comes from the fact that number of lanthanide transitions increases 
with energy, with the result that the light emerging at at peak emission is red. 

As mentioned earlier, unless the collapse is prompt, electron and neutrino captures deplete 
the neutrons in ejected matter, and this processed part of the ejecta is  
lanthanide poor, more transparent and, in particular, transparent to shorter wavelength light.  
Because the light emerges sooner, Metzger and Fern\'andez\cite{Metzger2014} and 
Kasen et al.\cite{kfm15} predicted early blue emission, 
lasting about two days in mergers with hypermassive neutron stars, 
and evolving to red and infrared. 

\vskip 0.3cm
\noindent{\sl The GW170817 kilonova} 

From ultraviolet through infrared, observations of the GW170817 remnant 
(labeled the astronomical transient AT2017gfo) strikingly confirmed 
the behavior of a radioactively-powered kilonova (see, e.g., \cite{Metzger2019} and references therein).  
A spectrum initially peaked in 
ultraviolet progressed through blue to red over about three days and then to 
infrared \cite{Evans2017,McCully2017,nicholl17}, supporting models whose 
ejecta had a range of neutron fractions.  Subsequent observations strengthen the 
case:  Watson et al.\cite{Watson2019} identify spectral lines of Sr, a light 
neutron-capture element, providing direct evidence that the early blue light 
and its rapid fading is associated with decays in a lanthanide-poor component. 
And the apparent match to decays in lanthanide-enriched matter of the slower 
progression through longer wavelengths was strengthened by late-time infrared observations 
at 43 and 74 days.\cite{Kasliwal2019} In that time, the measured infrared luminosity dropped by a 
factor of 6, a decline consistent with decays from 
a small set of nuclides with half-lives of roughly 14 days.  There are no light 
radioactive nuclei with decay times close to that, and it matches some of the more 
abundant heavy $r$-process elements (e.g., $^{143}$Pr, 13.6 d;$^{56}$Eu, 15.2d). 

Because merger simulations that produce lanthanides ordinarily yield elements up to 
and beyond the third peak, it is likely that the GW170817 merger synthesized 
the full range of $r$-process elements.  Given the estimated $0.03$ to $0.06\, M_\odot$ of 
$r$-process ejecta in this event, there is little doubt that mergers involving 
neutron stars contribute a major part of the universe's $r$-process matter. 
There is, however, evidence that they are not the only $r$-process site.  An example 
is a high observed abundance of Eu in the early universe -- too high for the 
early-time merger rate inferred from the event rate of observed SGRBs.\cite{Cote2018}

\section{Neutron star radius and tidal deformability}
\label{s:radius}

We turn now to measuring neutron star radii.  Our emphasis is on extracting radii from gravitational 
wave observations of binary inspiral and coalescence, but we also briefly review 
measurements of radii from recent electromagnetic observations.  

The gravitational wave measurements rely on the way tides alter the inspiral waveform.  
Tidal distortion of the stars increases the rate of inspiral, and it leads to an earlier coalescence -- 
and thus to a lower wave frequency at the time of maximum amplitude, approximately the time of contact. 
To estimate these effects and for our later calculation of the Hubble constant, 
we will need the quadrupole formulas for the wave amplitude $h$ and the rate $\dot E$ 
at which the orbit of a binary system loses energy to gravitational waves (see, for example, 
Wald\cite{wald} or Schutz\cite{schutz09}).  At lowest  post-Newtonian order, $h$ is linear and 
$\dot E$ quadratic in the Newtonian quadrupole moment tensor $Q_{ij}$: 
In Cartesian coordinates with origin at the Newtonian center of mass, the 
tensor has components
\be
Q_{ij} := \int dV\rho \left(x_i x_j - \frac13\delta_{ij} r^2\right). 
\label{e:Qij}\ee
The wave amplitude is an average magnitude of the perturbed metric $h_{\mu\nu}$. 
With the $z$ axis along the direction of propagation of the wave seen by an observer at 
large $r$, its nonzero components (in the Lorenz or transverse-tracefree gauge) are 
\[
    h_+:= h_{xx} = -h_{yy} = \frac1r(\ddot Q_{xx}- \ddot Q_{yy}), \qquad h_\times := h_{xy} = \frac2r \ddot Q_{xy};  
\]
the amplitude is 
\[ 
  h = \sqrt{h_+^2+h_\times^2}.
\]
For a periodic source, $h$ is twice the fractional change in the displacement of two free masses.  
The rate of energy loss has the form   
\be
 \dot E_{GW} = - \frac15\frac{c^5}G \dddot Q_{ij}\dddot Q^{ij}.
\label{e:Edotq1}\ee

This is enough to let us quickly find the amplitude and rate of energy loss from a binary system.
Let $\bm d(t)$ be a connecting vector joining stars of mass $m_A$ and $m_B$ in circular orbit.  We denote by $M=m_A+m_B$ and $\mu= m_Am_B/M$ the total and reduced masses of the system.  
From its definition \eqref{e:Qij}, the quadrupole tensor is 
\be
   Q_{ij} = \mu \left(d_i d_j-\frac13\delta_{ij}d^2\right).
\label{e:Iij}\ee
The spherical symmetry of each star implies that the quadrupole moment about its own 
center of mass vanishes. Choosing unit vectors $\hat{\bm e}_1$ and $\hat{\bm e}_2$ in the plane of the orbit, we can write 
$\bm d = d(\hat{\bm e}_1\cos\Omega t + \hat{\bm e}_2 \sin\Omega t)$.  Then  
\[
  Q_{11} = \frac12\mu d^2\left(\cos{2\Omega t}+\frac13\right), \  
  Q_{22} = -\frac12\mu d^2\left(\cos{2\Omega t} +\frac13\right), \ 
  Q_{12} = \frac12 \mu d^2 \sin 2\Omega t, 
\]    
implying a wave frequency 
\be
\omega=2\Omega, 
\ee        
amplitude averaged over a solid angle, 
\be
   \bar h = \sqrt{\langle h^2 \rangle} = \sqrt{\frac 85}\frac{G}{c^4}\frac1{r}\mu d^2 \Omega^2 ,
\label{e:amplitude}\ee
and a rate of energy loss given by 
\be
  \dot E_{GW} = -\frac{c^3}G \langle {\dot h}^2\rangle r^2 = -\frac{32}5\frac G{c^5}\mu^2 d^4 \Omega^6. 
\label{e:Edotq2}\ee

Because the frequency of the wave is directly measurable, while the 
binary's separation is not, it is helpful to use Kepler's law, 
$d^3 = GM/\Omega^2$, to replace $d$ by the GW frequency $f_{GW}$  
in these expressions for $h$ and $\dot E$.  
Writing $\Omega=\omega/2=\pi f_{GW}$, setting $G=c=1$, and omitting numerical factors, we have
\footnote{With $G$, $c$, and numerical factors restored, the expression for $\dot E$ is 
\[
  \dot E = -\frac{32}5\frac{G^{7/3}}{c^5} (\pi M_{\textrm{chirp}}f_{GW})^{10/3}.
\]}
\bsube\be
 h \sim \frac{M_\mathrm{chirp}^{5/3}}r f_{GW}^{2/3}, 
\label{e:h}\ee
\be
\dot E_{GW} \sim (M_\mathrm{chirp} f_{GW})^{10/3},
\label{e:Edotq3}\ee\esube
where the {\sl chirp mass} $M_\mathrm{chirp}$ is
\be
{M}_\mathrm{chirp}=\frac{\left(m_A m_B \right)^{3/5}}{M^{1/5}}.
\label{eq:Mchirp}
\ee 
Because only the chirp mass appears in these expressions, only the 
chirp mass is observable from the inspiral waveform at lowest post-Newtonian order 
(and neglecting tides). 
With the mass ratio $q=m_1/m_2$ entering at higher post-Newtonian order, its effect 
on the dynamics and waveform are more pronounced in the last part of the inspiral\cite{Creighton2011,Blanchet2014}.

To find the frequency $\omega$ as a function  of time one can integrate Eq.~\eqref{e:Edotq3} 
with $E$ having its Newtonian value, 
\be
 E=-Gm_Am_B/2d=-2^{-5/3}G^{2/3}M_\textrm{chirp}^{5/3}\omega^{2/3}, 
\label{e:Eomega}\ee 
to find 
\be
   \omega(t) = \frac{5^{3/8}}4 \left(\frac{GM_\textrm{chirp}}{c^3}\right)^{-5/8}
				(t_{\rm merger} -t)^{-3/8}.
\label{e:omega}\ee  
The total phase is then $\phi=\int \omega dt$, or 
\be
   \phi(t) = -2 \left(\frac{5GM_{\rm chirp}}{c^3}\right)^{-5 / 8} (t_{\rm merger} -t)^{5/8}+\phi_0, 
\label{e:phi}\ee  
where $t_{\rm merger}$ is the time at merger and $\phi_0$ is an initial phase.

\subsection{Radius from inspiral waveform}

A principal goal of gravitational-wave astronomy is to constrain the neutron star 
equation of state by looking at the imprint of tides on gravitational waves emitted in 
the inspiral and coalescence of NS-NS (and perhaps NS-BH) binaries. 
The tidally induced departure of the inspiral waveform from that of point particles (in a 
post-Newtonian framework) or of a spinless binary black hole system  
increases with the stiffness of the equation 
of state.  A stiffer equation of state, where the pressure increases rapidly 
with density, yields stars with larger radii and larger tidal effects on the 
waveform, governed by the star's {\it tidal deformablity}.
Following early work by Kochanek \cite{Kochanek} and Lai \& Wiseman\cite{LW} 
a growing research effort has explored the effects of tides on binary inspiral and coalescence 
and on extracting physical parameters from the observed waveforms: See, e.g., the Ligo/VIRGO
analyis of neutron star properties from GW170817\cite{dns_properties18} for an extensive list of 
references.

As a neutron star binary loses energy to gravitational waves, the orbit shrinks, giving a 
chirp-like increase in both the amplitude and frequency of the emitted waves.  
In the late inspiral, tides alter the waveform in two ways at the same order in the ratio 
$R/d$ of the stars' radius to the distance between them:   \\
(1) Orbital energy is lost directly to work done in raising tides. \\
(2) Indirectly, by distorting the stars, tides increase the system's quadrupole moment 
and thereby increase the rate at which energy is radiated in gravitational waves.

Each of these sinks of orbital energy $E$ accelerates the system's inspiral. 
After estimating the height of tides, we consider each in turn, 
finding that each sink increases $|\dot E|$ by a factor of order $R^5/d^5$.  
 
Consider a system of two stars, each of mass $m$ and radius $R$, in a circular orbit with 
separation $d$.   
With no tides, $d$ decreases as the system radiates gravitational waves, the total energy of 
its orbit decreasing at the rate $\dot E = \dot E_{\rm GW}$.  
Because the orbital energy of the binary system is $E = -Gm^2/2d$, the rate 
at which it loses energy to gravitational waves is related to $\dot d$ by 
\be
	\dot E_{\rm GW} /E = \dot d/d.
\label{e:Edotddot} \ee  
To estimate the tidal effects, we begin with a quick estimate of the height of tides.  
Fig.~\ref{f:tide} shows the elliptical (quadrupole) distortion of one of the two stars for a tide of 
height $h\ll R\ll d$. The initial spherical star is represented by a blue and purple circle with labeled radius.
 
\begin{figure}
\centerline{	\includegraphics[width=.85\textwidth]{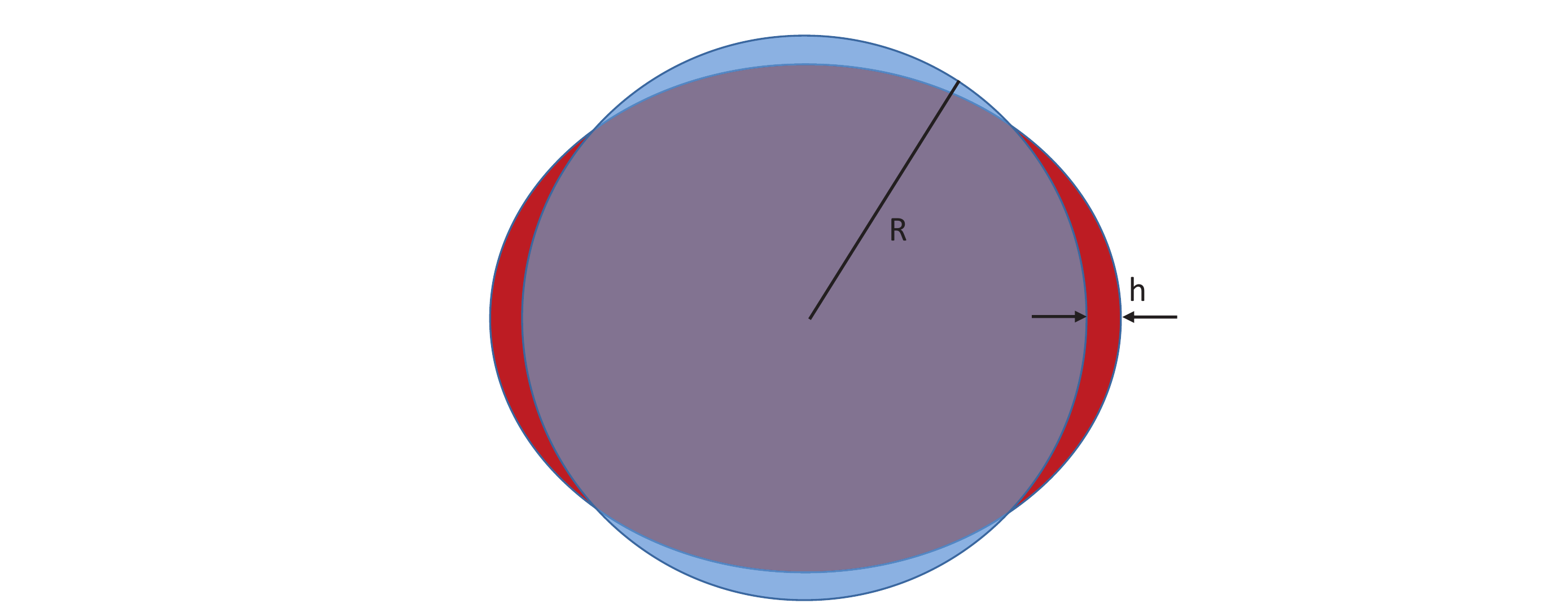}}
	\caption{ At linear order in the height {\texttt h} of a tidal bulge, 
the distortion is quadrupole.
For a uniform-density fluid, its effect is equivalent to moving the fluid in the blue arc-shaped 
regions at top and bottom to fill the red arcs at left and right.}
\label{f:tide}\end{figure}
For an incompressible fluid, the distortion is equivalent 
to moving the fluid in the arc-shaped regions of mass $\delta m$ on top and bottom 
to the arcs on left and right.  This involves moving the mass $\delta m$ a distance 
of order $R$ in the tidal field and raising it a height $\texttt h$ in the spherical 
gravitational field its own star.

The work needed to raise the mass $\delta m$ is of order $\delta m\, g\, \texttt h$ and is 
equal in magnitude to the negative work done by the 
tidal field (in a rotating frame) in moving the fluid a distance of order $R$.  
The tidal force on $\delta m$ is of order $\displaystyle \frac{Gm R}{d^3}\delta m$. 
Equating the work done by the tidal field to the work needed to raise the tide, 
we have \vspace{-1,5mm}
\be
   Gm\frac{R^2}{d^3}\delta m  \sim \delta m g\, \texttt h = \delta m \frac{Gm}{R^2}\texttt h,  
\label{e:work}\ee
implying \vspace{-1,5mm}
\be
  \texttt h \sim \frac{R^4}{d^3}.
\ee

\noindent (1)  {\sl Energy loss from work done to raise tide.}\\
Because each arc extends over an area of order $R^2$, the mass in each arc is of order 
$\delta m\sim m\, \texttt h/R$. 
From our expression \ref{e:work} for the work done to raise the tide,
the rate which the binary loses energy to tidal distortion is then
\[
   \dot E_{\rm tide} \sim \frac d{dt} \left(\delta m \frac{Gm}{R^2}\texttt h\right) 
		     \sim \frac d{dt}\left(Gm^2\frac{R^5}{d^6}\right) 
		     \sim Gm^2\frac{R^5}{d^6} \frac{\dot d}d\ .
\]
 \vspace{-1.5mm}
Eq.~\eqref{e:Edotddot} now gives the change due to tides in the rate of energy loss, 
\be
    \frac{\dot E_{\rm tide}}{\dot E_{\rm GW}} \sim \frac{R^5}{d^5}.
\label{e:Edot_tide}\ee  

\noindent(2) {\sl Energy loss from enhanced quadrupole radiation.}

Tides increase the quadrupole radiation of the binary because the tidal bulge of each 
star about its center of mass corotates in line with the stars. In Eq.~\eqref{e:Iij},
the contribution to $Q_{ij}$ of each star's bulge about the star's center of mass  
then has frequency $\Omega$.  The tidal contribution to the system's quadrupole moment tensor is the quadrupole moment of the arcs of Fig. \ref{f:tide}, 
with signs representing the difference between the deformed and spherical configurations; its 
its eigenvalues have magnitudes $\delta Q\sim \delta m\ R^2$.  From the quadrupole formula~\eqref{e:Edotq1}, the enhancement $\delta\dot E_{GW}$ in energy loss is then  \vspace{-1.5mm}
\be
 \frac{\delta \dot E_{GW}}{\dot E_{GW}} \sim \frac{\delta Q}{Q}
                \sim \frac{\delta m\,R^2}{m\,\,d^2} \sim \frac{R^5}{d^5},
\label{e:Edot_GW}\ee
identical, up to a dimensionless numerical factor, with that of Eq.~\eqref{e:Edot_tide}.

The $(R/d)^5$ dependence means that tides are primarily important in late inspiral\cite{Damour2010,Blanchet2014}; phase accumulated from the earlier inspiral, however, induces a time shift that corresponds to a larger phase difference in the later high-frequency cycles.\,\cite{Read2013}

One can measure the neutron star radius by observing the departure of the waveform from that of an 
inspiraling spinless binary black hole system.  For the same mass, stars with larger radii 
have greater tidal distortion, leading to more rapid loss of orbital energy and to earlier merger. 
Because of this tidally enhanced energy loss, fewer cycles are needed to decrease the orbital 
separation and increase the gravitational wave frequency, implying a smaller total phase  
as a function of frequency.  Similarly, a smaller gravitational wave energy accompanies a given 
increase in orbital frequency, implying a smaller wave amplitude as a function of frequency. 

Fig.~\ref{f:phase} shows the decrease in total phase with increasing neutron star radius. 
Current ground-base detectors are sensitive to frequencies from a few tens of Hz to a few kHz 
(see, e.g.~\cite{Abbott2019}), corresponding to the last few hundred cycles of NS-NS inspiral over 
a time of about 2 minutes. Most of the phase correction is in the later part of this interval; 
for a rough guess of the tidal correction, we take $f_{GW} = 1$ kHz, corresponding to 
$d = 33$ km when $M=2.7M_\odot$, and write as the tidal correction to the last $100$ cycles as 
$\Delta n \sim (R^5/d^5) 100 \sim 1$ cycle for $R= 13$ km, or $\phi_{\rm tide}\sim 6\, R_{13}^5$ rad.

For a more careful estimate, we can use the $R^5/d^5$ tidal correction to $\dot E$ and the Newtonian 
relation $dE/E \propto d\omega/\omega$ to approximate the tidal contribution to 
$\dot\omega$ by 
\[
\dot\omega_{\rm tide} \propto \dot\omega \frac{R^5}{d^5}, 
\] 
which yields a tidally induced change in phase 
\be
   \phi_{\rm tide} \propto \frac{R^5}{(M_{\rm chirp}M)^{5/3}}(\omega_f^{5/3} - \omega_i^{5/3}). 
\label{e:phase1}\ee 
When $\omega_f\gg\omega_i$, we can use Kepler's law to write this in terms of 
the geometric mean of the initial and final distances $d_i$ and $d_f$: 
\[
  \phi_{\rm tide} \sim K \left(\frac R{\sqrt{d_i d_f}}\right)^5,
\]
with $K$ a numerical factor of order unity. Using $d_i=150$ km and $d_f=21$ km, corresponding to initial and 
final frequencies 100 Hz and 2000 Hz, we have $\phi_{\rm tide}\sim 2 K$ rad, and the numerical simulations associated with Fig.~\ref{f:phase} below give $K\sim 10$.
The tidal departure of numerical waveforms in Fig.~\ref{f:phase} are consistent with this until 
estimate until the last few cycles, where tidal disruption begins and $\phi_{\rm tide}$ rapidly increases. 
 
\begin{figure}[h!]
\centerline{\includegraphics[width=.7\textwidth]{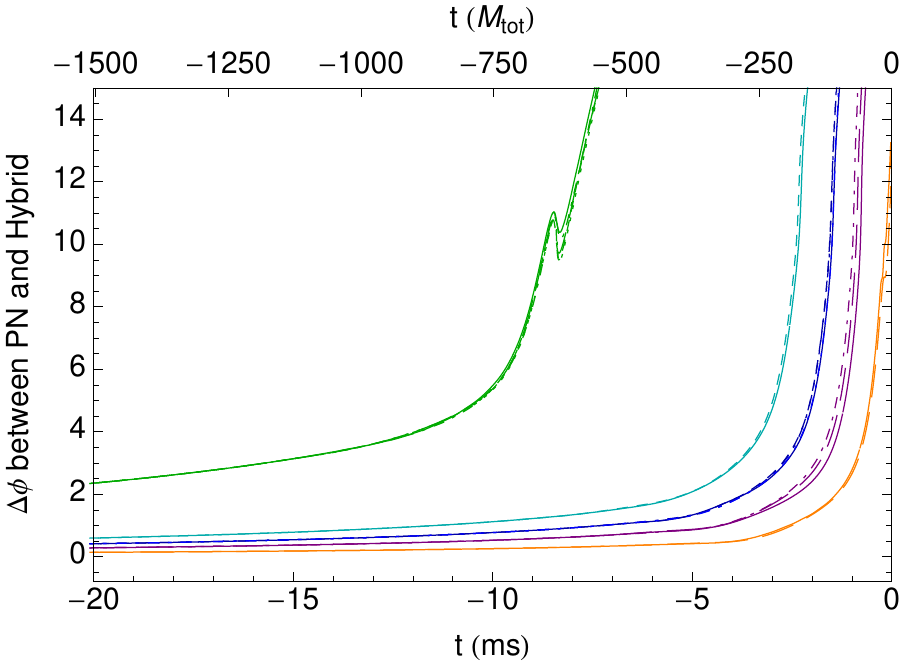}}
\caption{The tidally induced departure in phase of NS-NS inspiral from BH-BH inspiral 
(or point-particle inspiral in a post-Newtonian approximation), plotted against time before coalescence.
The stars each have mass $m = 1.35 M_\odot$.   
From highest to lowest, the lines show the phase departure for models with 
stellar radii 15.2 km, 12.3 km, 11.6 km, 11.0 km, and 10.3 km, corresponding to 
EOS 2H, H, HB, B, and Bss in Read et al.\cite{Read2013} Numerical integration is begun at 
200\,Hz, with a post-Newtonian waveform used at lower frequencies. This is Fig.~10 of Read et al.\cite{Read2013}}
\label{f:phase}\end{figure}

The tidal reduction in amplitude over a given frequency integral is shown in plots of the 
Fourier transform $\widetilde h(f)$ of $h(t)$.  Fig.~\ref{f:amplitude} shows the decrease in 
$|\widetilde h(f)|$, weighted by $f^{1/2}$, with increasing neutron star radius for a 
slightly narrower range of models than that of Fig.~\ref{f:phase}.  
The monotonically decreasing part of each curve describes the waveform during the inspiral, 
with the first minimum approximately at merger.  
Peaks seen after merger are are related to oscillation modes and are described in Sect.~\ref{post} below.     
\begin{figure}[h!]
\centerline{\includegraphics[width=.6\textwidth]{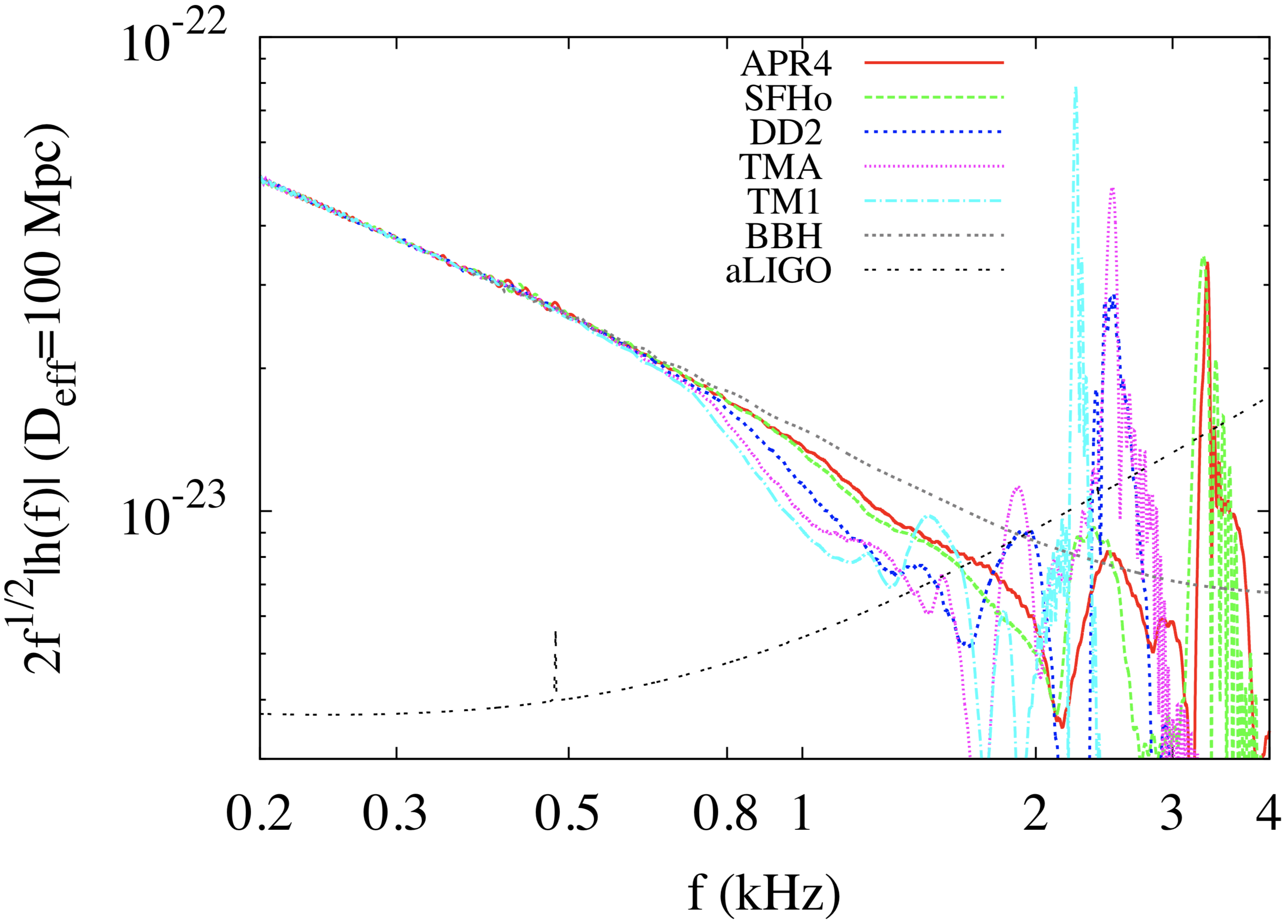}}
\caption{The inspiral gravitational wave amplitude as a function of frequency for two neutron stars 
each of mass $m = 1.35 M_\odot$.  
From highest to lowest, the lines correspond to models 
with stellar radii 11.1 km, 11.9 km, 13.2 km, 13.85 km, and 14.5 km. With this order, 
the models are based on EOS APR4, SFHo, DD2, TMA, and TM1, as listed in in Hotokezaka 
et al.\cite{hkss16}; this is Fig.~3 of that reference.}
\label{f:amplitude}\end{figure}    

We have estimated tidal effects as a function of neutron star radius, with corrections proportional to $R^5$.  In post-Newtonian waveforms, the departure from point-particle inspiral depends on the {\sl tidal deformability}, $\Lambda$, not explicitly on the radius, and the $R^5$ dependence is only 
approximate. We briefly outline the definition and role of $\Lambda$ in a general relativistic 
context.  

In the rest frame of one of the neutron stars, the quadrupolar tidal field sourced by the companion is $ \mathcal{E}_{i j}=R_{t i t j}$, where $
R_{\mu \alpha \nu \beta}$ is the Riemann tensor of the spacetime describing the companion only.
The neutron star responds to the tidal disturbance by its companion, by adjusting its internal structure to a new equilibrium configuration. At asymptotically large distances from its center, this adjustment enters in the form of multipole moments 
$$ 
\frac{1+g_{t t}}{2}=\frac G{c^2}\left[\frac{M}{r}+\frac{\left(3 n^{i} n^{j}-\delta^{i j}\right) Q_{i j}}{2 r^{3}}+\mathcal{O}\left(r^{-4}\right)-\frac{1}{2} n^{i} n^{j} \mathcal{E}_{i j} r^{2}+\mathcal{O}\left(r^{3}\right)\right],
$$
where $Q_{i j}$ is the neutron star's mass quadrupole moment tensor and \(n^{i}=x^{i} / r\) is a unit radial vector of the chart $\{ x^i\}$.    

 The neutron star's response to the tidal perturbation can be described in terms of excitations of its oscillation modes, which are either resonantly excited 
(when the tidal forcing frequency coincides with the mode frequency) or are otherwise adiabatically driven. For example, the tidally induced quadrupole moment $Q_{ij}$ is a sum of contributions from all quadrupolar $\ell=2$ modes.  Because the higher multipole moments of the tidal field fall off as higher powers of $R/d$, the $\ell>2$ contribution to observed inspiral waves can be neglected. 

When the orbital separation is much larger than the neutron star radius, the induced quadrupole moment is linearly proportional to the tidal field:
\begin{equation}
Q_{i j}^{\text {adiab }}=-\lambda \mathcal{E}_{ij},
\end{equation}
where $\lambda$ is the \textit{tidal deformability} parameter, related to the\textit{ tidal Love number} $k_2$ and the neutron star radius by \(\lambda=2 /(3 G) k_{2} R^{5}\). It is customary to define a dimensionless tidal deformability,
$$
\Lambda=\frac{ c^{10}}{G^{4} M^{5}}\lambda,
$$
which can be computed by solving a second-order ordinary differential equation, in addition to the well-known TOV equations for neutron star structure (see \cite{2019JPhG...46l3002G,2020arXiv200402527D,2020arXiv200406419B} for recent reviews).

In the frequency domain, the GW signal for a binary neutron star coalescence is\cite{PhysRevD.49.2658} 
$$
\widetilde{h}(f)=\mathcal{A} f_{\rm GW}^{-7 / 6} \exp \left[\mathrm{i}\left(\phi_{\text {point-mass }}+\phi_{\rm tide}\right)\right],
$$
where the amplitude $\mathcal{A}$ includes both point-mass and matter effects.  The tidal contribution to the frequency-domain GW phasing is\cite{PhysRevD.77.021502} 
\begin{equation}
\phi_{\rm tide}
 =-\frac{117}{256}\, \widetilde\Lambda 
	\left(\frac{\pi G}{c^3} \frac{M^2}{M_{\rm chirp}} f_{\rm GW}\right)^{5/3},
\end{equation}
with the dependence on mass and frequency we found in Eq.~\eqref{e:phase1}.  
The phase correction depends on the \textit{effective} (or weighted average)
tidal deformability
\begin{equation}
\widetilde\Lambda=\frac{16 c^{10}}{13 G^{4}} \frac{\left(m_1+12 m_2\right) m_1^{4} \Lambda_{1}+\left(m_{2}+12 m_{1}\right) m_{2}^{4} \Lambda_{2}}{M^{5}}
\end{equation}
which  can be constrained  through GW observations (for equal-mass binaries, $\tilde \Lambda = \Lambda_1 = \Lambda_2)$.
More elaborate models have been developed for matching theoretical waveforms of BNS mergers to GW observations (see\cite{2019JPhG...46l3002G,2020arXiv200203863R} and references therein).  
An accurate description in the high-frequency regime is obtained through the effective-one-body (EOB)\ formalism\cite{1999PhRvD..59h4006B,Damour2010}, where the relative motion of the two stars is equivalent to the motion of a particle of mass equal to the reduced mass $\mu=m_1m_2/M$ in an effective potential. 
 
The EOB two-body Hamiltonian for nonspinning binaries can be expanded as
\begin{eqnarray}
H_{\mathrm{EOB}} \simeq M c^{2}+\frac{\mu}{2} p^{2}+\frac{\mu}{2}\left(-\frac{2 G M}{c^{2} d^{2}}+\ldots-\frac{\kappa^{T}}{d^{5}}\right),
\end{eqnarray}
where $p$ is momentum. The constant $\kappa^T$ encodes the effect of quadrupolar tidal interactions at leading order in $R/d$ and in a post-Newtonian expansion. It is equal to  $\kappa^T = \kappa_1 +\kappa_2$, where
 $\kappa_{i}$ $(i=1,2)$  are the quadrupole tidal polarizability coupling constants for the individual stars, in a multipolar expansion of the tidal potential (see \cite{2020arXiv200203863R} for a recent review) and are equal to a function of mass times the corresponding tidal deformability $\Lambda_{i}$. Furthermore, phenomenological tidal models can be constructed by fitting EOB models to numerical relativity simulations\cite{CORE8,2018PhRvD..97d4044K,Lackey2019}.

   The chirp mass  is measured with high accuracy from the inspiral signal and for GW170817 it was obtained as\cite{Abbott2017,Abbott2019} $M_{\rm chirp} = 1.186(1)M_\odot$. With the current sensitivity, the total binary
mass of sources at a few tens of Mpc can be obtained with an accuracy of
order
one percent~\cite{Rodriguez2014,Farr2016,Abbott2017}.
For GW170817 the total mass was $ M_\mathrm{tot}=2.74^{+0.04}_{-0.01}~M_\odot$. 
By comparison, the mass ratio $q$ was poorly constrained to be between 0.7 and 1. 
The dimensionless tidal deformability was initially constrained to be $\tilde \Lambda < 800$ 
at the 90\% confidence level, assuming a uniform prior\cite{Abbott2017},\ which corresponds to $\kappa^{\rm T}<150$. A subsequent improved analysis\cite{Abbott2019} gave a 90\% highest posterior density interval of \(\widetilde{\Lambda}=300_{-230}^{+420}\). A somewhat different range for $\tilde \Lambda$\ was  recently obtained with other numerical-relativity calibrated waveforms\cite{2019arXiv191008971N}.    
   
   Constraints on $\Lambda$ can be translated to constraints on radii. For example, for $1.4\,M_\odot$ models, the empirical relation
\begin{equation}
\Lambda_{1.4}=2.88 \times 10^{-6}(R_{1.4} / \mathrm{km})^{7.5},
\label{annala}
\end{equation}
was found\cite{2018PhRvL.120q2703A}, leading to $R_{1.4}<13.6$km for $\Lambda_{1.4}<800$, taking into account the error bars in (\ref{annala}). The effective tidal deformability $\tilde \Lambda$ was also shown to correlate well with the radius of the primary star $R(m_1)$, for  a fixed chirp mass\cite{Raithel2018}. The initial analysis of the GW170817 merger\cite{TheLIGOScientificCollaboration2018} resulted in the constraint of $R=11.9_{-1.4}^{+1.4}$km (for both stars involved in the merger). 
A large number of other estimates  of NS
radii based on the observation of GW170817 (or in combination with multimessenger
and/or experimental constraints) have appeared, see e.g. \cite{2019EPJA...55...97T,De2018,Most2018,2018arXiv181110929M,Fattoyev2018,Radice2018a,2020NatAs.tmp...42C,2020arXiv200304880L}, and references in the review articles.\cite{2019PrPNP.10903714B,2019JPhG...46l3002G,2019EPJA...55...80R,2020arXiv200203863R,2020arXiv200402527D,2020arXiv200406419B}. 

Setting EOS constraints through multiple detections has been considered in\cite{2013PhRvL.111g1101D,2015PhRvD..92j4008C,2015PhRvD..91d3002L,2019PhRvD.100j3009H,2020PhRvD.101d4019C}. Besides  EOS  constraints, EOS-insensitive relations are  also useful in
extracting source  properties\cite{2019PhRvD..99h3016C}. 

\subsection{From electromagnetic observations} 

With varying accuracy and model dependence, a number of observational methods, 
have been used to measure neutron star radii (see, for example, summaries 
by Lattimer\cite{Lattimer2017} and by Oz\"el and Freire\cite{ozel16}).  

Particularly promising is the recently launched NICER instrument,  
which looks at X-rays from pulsars.  Charged particles that spiral around 
closed field lines of the rapidly rotating magnetic field collide with the surface 
at magnetic poles, creating X-ray emitting hot spots.  By accurately modeling the 
images of a rotating star with regions of varying temperature, the NICER project 
can match the observed periodic variation in X-ray intensity to obtain a best fit 
to the temperature distribution, mass and radius. Preliminary results\cite{NICER2019a} for the 
millisecond pulsar J0030+0451 give equatorial radius $12.71^{+1.14}_{-1.19}$ km 
and mass $1.34^{+0.15}_{-0.16}M_\odot$, associated with a preferred hot-region model.
Associated with an alternative temperature distribution\cite{NICER2019b} is an additional km uncertainty.
NICER anticipates observing several pulsars with eventual accuracy in radius measurement
to $\sim 0.5$ km.  

A second method uses {\sl quiescent} X-ray binaries, binary systems in which the 
neutron star accretes mass episodically from a companion and radiates steadily 
when it is not accreting.\cite{Lattimer2017,ozel16,bhog16,slb10}  
In this quiescent stage, assuming the radiation is thermal, the star behaves as 
a blackbody with intrinsic luminosity $A \sigma T^4$, with $A$ the star's surface area, 
and if the distance to the system is known, one can infer the radius.  
Two uncertainties are the amount of intervening interstellar matter to absorb 
the X-rays, and the composition of the outer atmosphere (by mistaking helium for 
hydrogen one underestimates the radius).  

A final related method uses the expansion of the neutron star's atmosphere 
in an X-ray burst, the explosive nuclear reaction at the star's surface that 
occurs when the amount of accreted matter reaches its critical mass. Again,
one extracts the star's radius from a blackbody temperature, in this case 
after the ejected atmosphere has settled back to the surface 
(see \cite{Strohmayer2006,Nattila2017} and references therein).

\section{NS maximum mass}
\label{s:mass}

\subsection{From post-merger} 

The maximum mass of a non-rotating neutron star, $M_{\rm max}$, represents the ultimate constraint on the EOS, in the sense that it refers to the star in which  the highest possible densities are sampled. From the observations of GW170817 and its electro-magnetic counterpart one can arrive at constraints on  $M_{\rm max}$, that are, however, dependent on the assumption on makes about the fate of the remnant. 
 
 By combining the total binary mass of GW170817 inferred from the GW signal with conservative upper limits on the energy in the GRB  and and in the ejecta from EM observations, a relatively short-lived remnant is favoured, setting an upper limit of \(M_{\max } \lesssim 2.17 M_{\odot}(90 \%) \)  \cite{Margalit2017a}.
Rezzolla {\sl et al.}\cite{rezzolla18} argue that the remnant survived for a longer time and collapsed  as a supermassive neutron star near the maximum mass supported by uniform rotation. Using an empirical relation between the maximum mass for nonrotating models and uniformly rotating models, they arrive at an upper limit of \(M_{\mathrm{max}}  \lesssim 2.16_{-0.15}^{+0.17}\, M_{\odot}\) (notice the large uncertainty).
A longer-lived remnant surrounded by a torus was also considered in \cite{kyoto17}
 and, in combination with the absence of optical counterparts from relativistic ejecta, the maximum mass was argued to be in the range of \(2.15-2.25 M_{\odot}\).

Under different assumptions (a BH is formed in a delayed collapse soon after the NSNS merger and the sGRB is triggered
by collimated, magnetically confined, helical jet and powered by a magnetized disk) one can write\cite{ruiz18} 
 \begin{equation} \beta M_{\mathrm{max}} \approx  M_{\rm max,rot}\, \lesssim \,M_{\mathrm{GW170817}} \approx 2.74 M_\odot\,\lesssim \, M_{\text {thresh }} \approx \alpha M_{\max },
\end{equation} 
with
\(\alpha \approx 1.3-1.7\) from numerical simulations\cite{Shibata2005,bbj13} and \(\beta \approx 1.2\) for microphysical EOS\cite{1987ApJ...314..594F,1994ApJ...422..227C,1994ApJ...424..823C,2016MNRAS.459..646B}.  This implies an upper limit of \(M_{\max } \lesssim 2.16 M_{\odot} \). A more conservative upper limit based on causality is obtained by considering the relations\cite{Koranda1997} 
\begin{equation}
M_{\max }^{\mathrm{}}=4.8\left(\frac{2 \times 10^{14} \mathrm{g}/\mathrm{cm}^{3}}{\epsilon_{m} / c^{2}}\right)^{1 / 2} M_{\odot} \ \ \mathrm{and} \ \ M_{\rm max,rot }=6.1\left(\frac{2 \times 10^{14} \mathrm{g} / \mathrm{cm}^{3}}{\epsilon_{m} / c^{2}}\right)^{1 / 2} M_{\odot},
\end{equation}
where $\epsilon_{m}$ is a matching 
energy density, above which the EOS\ is assumed to be at the causal limit\ (speed of sound equal to speed of light). Then, \(\beta \approx 1.27\) and  \(M_{\max } \lesssim 2.28 M_\odot\). Similar considerations were used in. \cite{Shunke2020}

A prompt collapse to a BH ($M_{\rm tot}>M_{\rm thres})$ is expected to be accompanied by systematically less massive and more neutron-rich ejecta,
resulting in a less luminous and redder kilonova than for the case of a delayed collapse\cite{Margalit2019} (see{\cite{2020PhRvD.101d4006A} for an alternative method to infer the threshold mass). In this case, one can use the empirical relation\cite{bbj13}
\begin{equation}
M_{\mathrm{thres}} \approx\left(-3.606 \frac{G M_{\mathrm{max}}}{c^{2} R_{1.6}}+2.38\right) M_{\mathrm{max}},
\label{kempir}
\end{equation}
where $R_{1.6}$ is the radius of a $1.6 M_\odot$ star, to directly constrain $M_{\rm max}$. Combining a large number of future observations is expected to tighten the constraints on  $M_{\rm max}$\cite{Margalit2019} and other EOS properties\cite{2020ApJ...888...12M}.
\vspace{-4mm}

\subsection{From binary systems} 

The three largest accurately measured neutron star masses have values close to 2.0$M_\odot$.
The stars, J1614-2230\cite{demorest10}, J0348+0432\cite{Antoniadis2013}, and, most recently 
J0740+6620\cite{Cromartie2020}, have, respectively, measured masses $1.97\pm 0.04M_\odot$, 
$2.01\pm 0.04$, and $2.14\pm 0.1 M_\odot$ (the uncertainties represent 1-$\sigma$ errors).  
Each is in a binary system, and each orbit 
is circular to one part in $10^6$.  Two of the systems are eclipsing binaries: 
Seen almost edge on, their orbital planes are nearly in the line of sight  
(angles $i$ between plane perpendicular to line of sight and plane of orbit are 
$89.17^\circ$ for J1614+2230 and $i=87.35^\circ$ for J0740+6620). 

Obscured by the complexity of the detailed analyses is 
an underlying simplicity that is exact for circular orbits in the line of sight:
For Newtonian binaries (and these are nearly Newtonian), two equations among the 
three unknowns, $m_1$, $m_2$ and the orbital radius $a$, are Kepler's third law, 
  $ a^3 = GM/\Omega^2$, 
and the expression for the orbital velocity $v_1 = r_1\Omega$ of the neutron star in 
terms of its distance $ r_1 = a\ m_2/M$ from the system's 
center of mass: $v_1 = \frac{m_2}M a\Omega$.
The period $P$ of the orbit is given by the periodic Doppler shift in the 
observed time interval between pulses, and the pulsar velocity $v_1$ relative 
to the system's center of mass can be measured from half the difference between 
the maximum blue- and red-shifts.  The third measurement needed to determine the 
three variables $m_1,m_2,a$ is the Shapiro time delay, (see, for example \cite{wald}), 
the delay in light travel time along a light trajectory that grazes the companion with a 
distance $r_0$ of closest approach.     
With $t$ proper time measured on Earth and $D$ the distance from the pulsar to the Earth, 
the delay is, to lowest order in $M/r_0$, $r_0/D$ and $r_0/a$, \vspace{-2mm}
\be
   \Delta t = \frac{2Gm_2}{c^3}\ln\left(8aD/r_0^2\right), 
\vspace{-2mm} \label{e:delay}\ee 
where $r_0$ is the distance of closest approach of the light ray to the center of 
the companion, and $D$ the distance of the binary system to the Earth.   
The distance $D$, however, is not accurately measured, 
and, in practice, one measures not the absolute time delay, but its change due to the change in $r_0$ as 
the pulsar orbits. Eq.~\eqref{e:delay} implies the difference is independent of $D$:  
 \be
   \Delta t(r_0')-\Delta t(r_0) = \frac{2Gm_2}{c^3}\ln\left(r_0^2/r_0'^2\right), 
\vspace{-2mm} \ee 
Because the ratio of impact parameters depends only on the observed angular pulsar positions, 
it can be directly measured.  The measured time delay now determines $m_2$, and the 
remaining two relations then give $a$ and the pulsar mass $m_1$.    

 Because the Shapiro time delay is small, of order $10^{-5}$ s, the analysis for the 
real system must include the comparable small corrections to a circular Newtonian orbit:  
post-Newtonian corrections as well as eccentricity and orbital inclination.  
As usual, orbital measurements rely on accurate timing of the extremely stable rotation of old neutron stars.
       
In the remaining system, J0348+0432, a white dwarf and neutron star orbit in an 
plane far from the line of sight.  In this case, the Shapiro time delay cannot be accurately measured;
instead, the authors obtain $m_2$ from the mass-radius relation for low-mass white dwarfs, 
estimating the radius and surface gravity from hydrogen Balmer lines at the surface of the dwarf.

\section{Neutron star EOS}
\label{s:eos}

 Above nuclear saturation density, $\rho_n = 2.7\times 10^{14} \rm g/cm^3$, and up to at least a few times 
nuclear density, the star consists primarily of neutrons with a small fraction of protons, electrons, 
and muons.  The primary uncertainty in its composition is in the star's dense core.  At high enough 
density, the Fermi energy of down quarks in compressed nucleons must exceed the 
rest mass of strange quarks, and a transition from nucleons to hyperons occurs; and 
at a presumably higher density, whose value in cold matter is similarly uncertain, the 
nucleons themselves dissolve, creating strange quark matter, comprising free up, down, and strange quarks.  
Should these critical densities be below the central density of the maximum-mass neutron star, 
the phase transitions to hyperons and/or strange quark matter will soften the equation of state 
making the star more compact and lowering the maximum mass.    

 Because old neutron stars are cold, with thermal energy far below the Fermi energy, their 
matter satisfies a one-parameter EOS of the form $P=P(\rho)$, where $\rho$ is the baryon mass density 
and $P$ the pressure of the star.  The EOS determines the one-parameter family of neutron stars, 
with the star's mass a function of its radius or central density, and the relation can be inverted: 
The $M(R)$ curve can be inverted to give $P(\rho)$.\cite{lindblom10}.  Although simultaneous measurements 
of mass and radius are currently restricted to stars whose mass is of order $1.4 M_\odot$, these,
together with causality and the maximum neutron star mass, substantially restrict the universe 
of candidate EOSs. 

   The radius of a 1.4$M_\odot$ neutron star is closely correlated with the pressure at about twice 
nuclear density.\cite{lp01}. For EOSs whose maximum masses range from 2.5\,$M_\odot$ to 
$2.0 \,M_\odot$, corresponding central densities range from about $2\times 10^{15}$ to $2.9\times 10^{15}$ g/cm$^3$.
The maximum mass is then approximately governed by the pressure at densities of 
order $7$-$8 \rho_n \sim 2\times 10^{15}$ g/cm$^3$.\cite{op09}.  The diagram below, patterned after 
\"Ozel \& Freire\cite{ozel16} portrays the approximate relation.    
\begin{figure}[h!]
\centerline{   \includegraphics[width = 6cm]{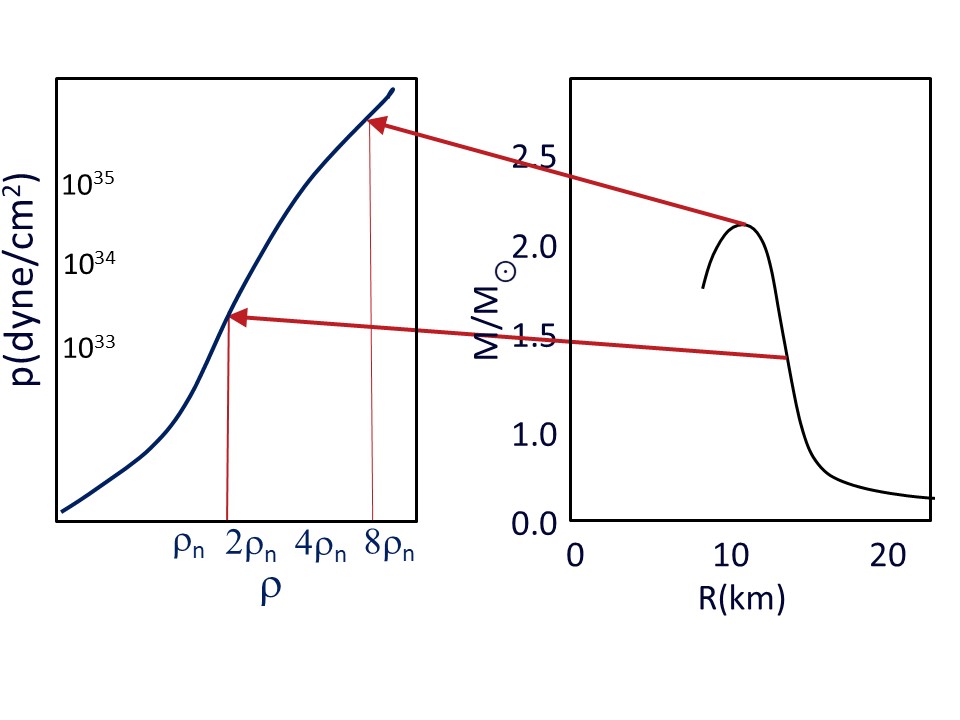}}
       \caption{Maximum mass and the radius at about 1.4$M_\odot$ roughly correspond to the parts of the 
EOS curve sketched here.  }
\label{fig:valley}\end{figure}

To systematize the observational constraints on the neutron star equation, Read {\sl et al.}\cite{rlof09} 
introduce a parameterized equation of state above nuclear density, specifying the pressure at three 
fiducial densities with linear interpolation of $\log p$ vs $\log\rho$.  Using these {\sl piecewise polytropic} 
models, subsequent authors have used GW and electromagnetic observations, causality, and 
constraints from nuclear theory to constrain the $p(\rho)$ 
curve.\cite{op09,lp16,dns_eos18,de18,cwi18,Lattimer2017}  An alternative spectral representation of the EOS due to Lindblom\cite{lindblom10,lindblom18} can give a more accurate map from observational constraints to the EOS\cite{cwi18} with a less obvious physical interpretation of the parameters in the spectral expansion.

\subsection{Constraints from causality and minimum radius} 

    \begin{figure*}
\begin{center}
\resizebox{1\textwidth}{!}{
\includegraphics{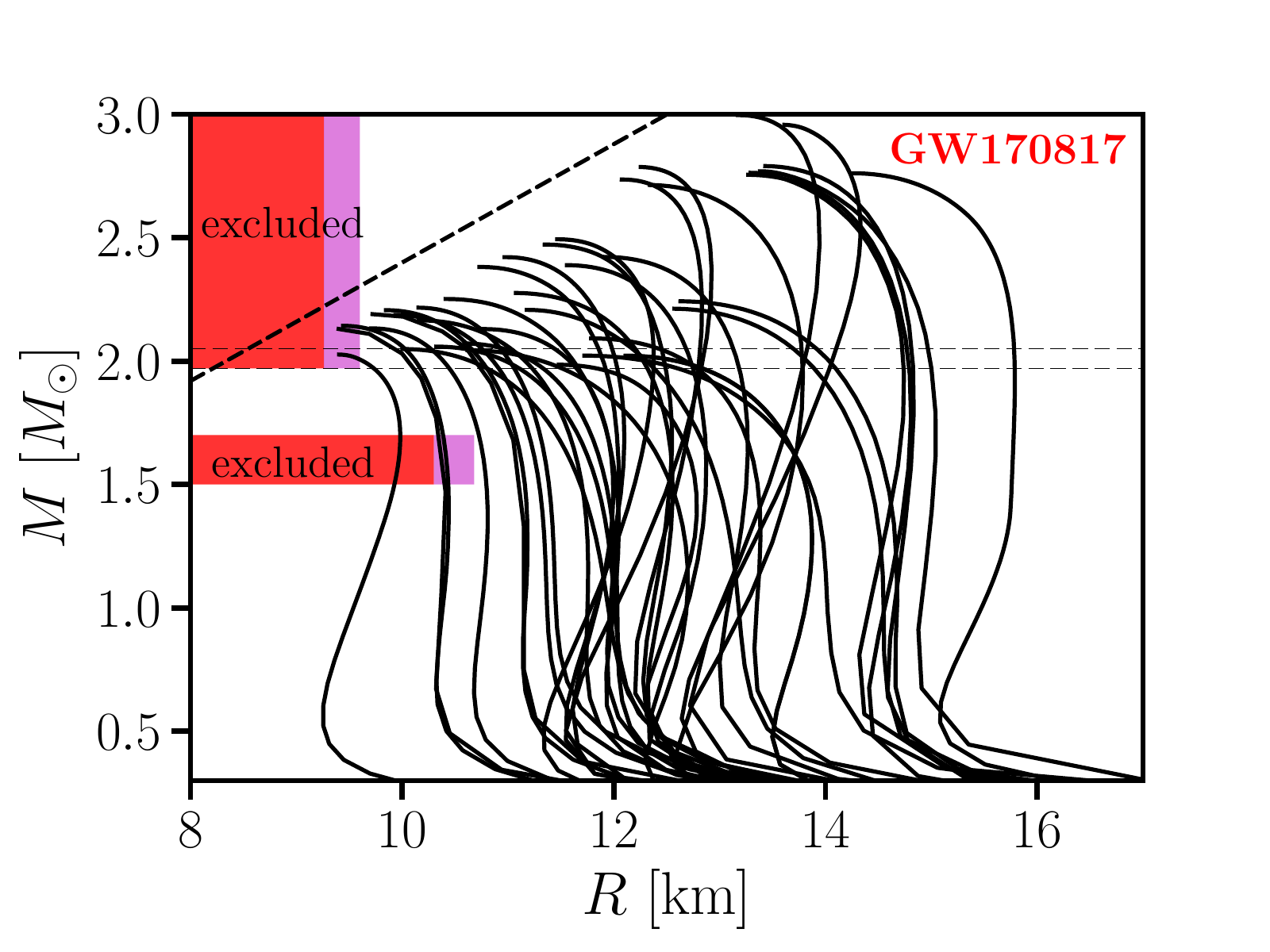}
\includegraphics{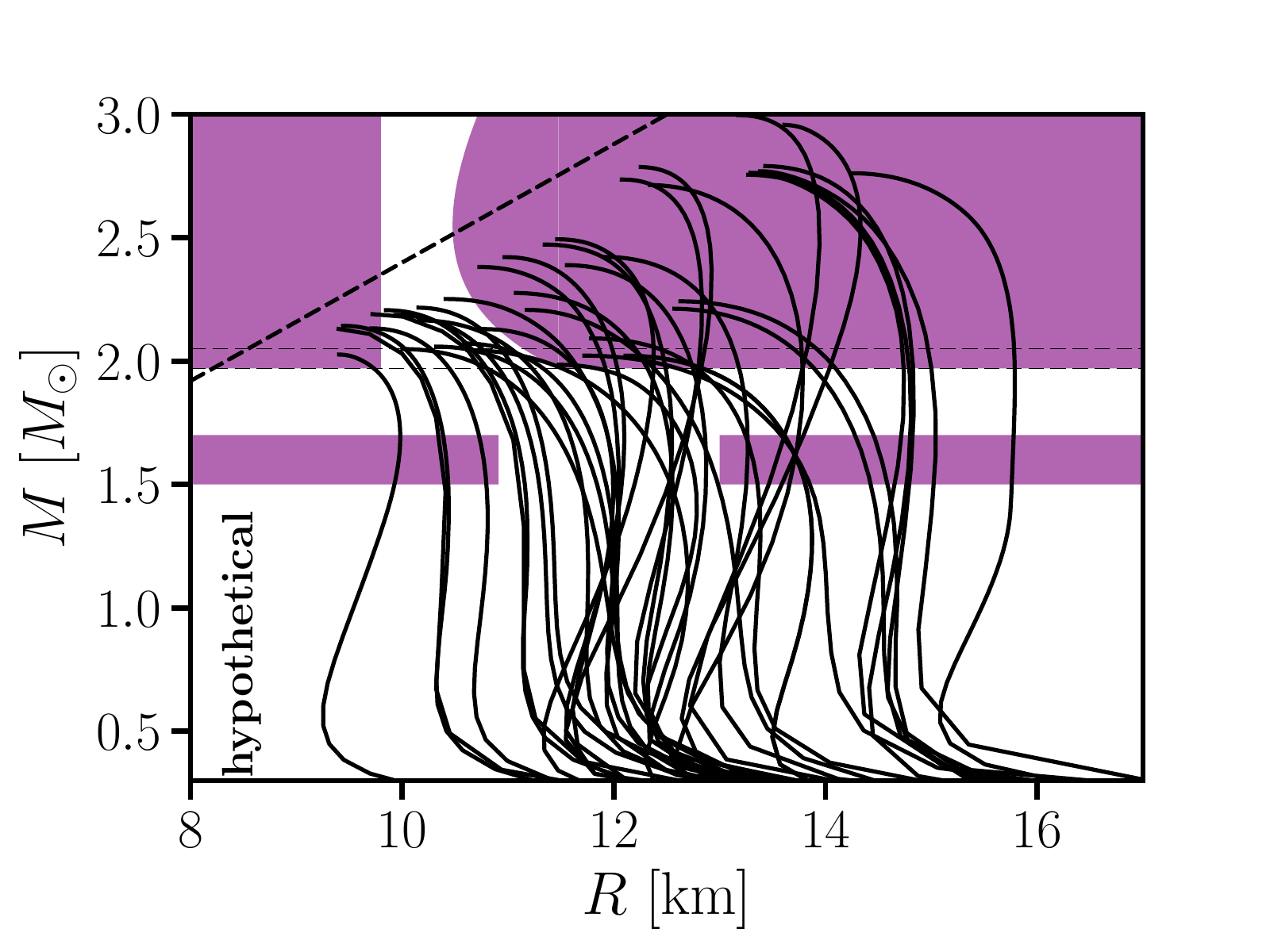}
}
\end{center}
  \caption{Left panel: Mass-radius relations of different EoSs with very
conservative (red area) and ``realistic'' (cyan area) constraints derived
from the measured total binary mass of GW170817 under the assumption of no
prompt BH formation of the merger remnant (see \cite{Bauswein2017} for details).
Horizontal lines display the limit by \cite{Antoniadis2013}. The thick dashed
line shows the causality limit Eq. (\ref{causemax}). Right panel: Hypothetical
exclusion regions (purple areas) from a delayed-collapse
event with $M_\mathrm{tot}=2.9~M_\odot$ and a prompt-collapse event with
$M_\mathrm{tot}=3.1~M_\odot$.
Figures from \cite{Bauswein2017}.}
  \label{fig:radcon}
\end{figure*}
 
Causality, in the form $v_{\rm sound}< c$, where $v_{\rm sound} = dp/d\epsilon$, with $\epsilon$ 
the energy density.%
\footnote{The propagation of signals along characteristics of relativistic fluid equation 
exceeds the speed of light unless this relation is satisfied for fluids obeying a one-parameter 
EOS or a for a stable star with a 2-parameter EOS of the form $p=p(\rho,s)$, with $s$ the 
specific entropy.\cite{vof}}  
significantly constrains the EOS above nuclear density.  Because the upper limit on mass 
appears to exceed 2.1$M_\odot$, the EOS must be stiff at high density, and causality 
then requires a minimum pressure above a few times nuclear density.     
Because stiffer EOSs yield stars with larger radii, a large upper mass limit then sets 
a lower limit on the radius of neutron star.  We now obtain that limit, beginning with 
the lower limit on the radius of the maximum mass star.    
  
Using the maximally soft EOS consistent with causality\cite{KSF97} and with a given $M_{\rm max}$, 
Haensel {et al.}\cite{Haensel1999} and Lattimer\cite{Lattimer2012} obtain 
\begin{equation} M_\mathrm{max}\leq\frac{1}{2.82}\frac{c^2\,R_\mathrm{max}}{G},
\label{causemax}
\end{equation}
where $R_\mathrm{max}$ is the radius of the maximum-mass nonrotating star.  This implies
that an EOS cannot become arbitrarily stiff (see also \cite{Koranda1997}).  There is a tight empirical relation\cite{bbj13}, analogous to Eq.~\eqref{kempir}, between
$k=M_\mathrm{thres}/M_\mathrm{max}$ and $\displaystyle C_\mathrm{max}=\frac{G\,M_\mathrm{max}}{c^2\,R_\mathrm{max}}$\,,
\begin{equation}\label{eq:mthr}
M_\mathrm{thres}=\left(  -3.38\frac{G\,M_\mathrm{max}}{c^2\,R_\mathrm{max}}+2.43
 \right)\,M_\mathrm{max}.
\end{equation}
Inserting the causality constraint Eq. (\ref{causemax}) in  Eq.~(\ref{eq:mthr}) results
in the constraint
\begin{equation}\label{eq:radcon}
M_\mathrm{thres}\leq0.436\frac{c^2\,R_\mathrm{max}}{G}, \ \ \ \mathrm{or} \ \ \ R_\mathrm{max} \geq 2.29\frac{G\ M_{\rm thres}}{c^2}.
\end{equation}
Hence, a given measurement or estimate of $M_\mathrm{thres}$ sets a lower
bound on $R_\mathrm{max}$. 

  As previously mentioned, electromagnetic observations of GW170817 give an ejecta mass between 0.03 and 0.06~$M_\odot$.  
Based on this mass range,
it was suggested in \cite{Bauswein2017} that the merger did not result in a prompt collapse, 
because direct
BH formation implies significantly reduced mass ejection (see e.g. Fig.~7
in \cite{Bauswein2013a}). This implies that the measured total
binary mass of GW170817 is smaller than the threshold binary mass $M_\mathrm{thres}$
for prompt BH formation and thus $M_\mathrm{thres}> 2.74^{+0.04}_{-0.01}~M_\odot$.
Using this condition in Eq.~(\ref{eq:radcon}) results in a lower limit on
$R_\mathrm{max}$.
The detailed calculation and error analysis in
\cite{Bauswein2017} yields $R_\mathrm{max}\geq9.60^{+0.14}_{-0.03}$~km.

Following
the same line of arguments, one can set a lower
limit on
the radius $R_{1.6}$ of a nonrotating 1.6\,$M_\odot$ NS. Replacing the EOS above 1.6\,$M_\odot$ with the causal limit EOS, yields 
\begin{equation}
M_{\max }=\frac{1}{3.10} \frac{c^{2} R_{1.6}}{G},
\label{cause1.6}
\end{equation}
and combining this with Eq. (\ref{kempir}) one obtains\cite{Bauswein2017}  $R_{1.6}\geq10.68^{+0.15}_{-0.04}$~km. The minimal set of assumptions,  Eq. (\ref{kempir}, \ref{cause1.6}) thus set an absolute
lower limit on NS radii that rules out very soft nuclear matter. 
The lower limits on $R_{\rm max}$ and $R_{1.6}$ are shown in the left panel of Fig.~\ref{fig:radcon} on top of a set
of mass-radius relations of a large sample of EOS.

It
is straightforward to convert the above radius constraints to a limit on the tidal
deformability. A limit of $R_{1.6}>10.7$~km corresponds to a lower bound
on the tidal deformability of a 1.4~$M_\odot$ NS of about $\Lambda_{1.4}>200$ (constraints on the tidal deformability that rely on the post-merger properties of GW170817, not all consistent which each other, were also derived in e.g.\cite{Radice2018,Coughlin2018,2019ApJ...876L..31K}).

   With future observations,  an event that is identified as a prompt collapse  will set an upper bound on
$M_\mathrm{thres}$, which constrains
the maximum mass and radii of nonrotating NSs from above. The resulting constraints
from a hypothetical future detection are shown in Fig.~\ref{fig:radcon} (right
panel).
A large number of detections of the inspiral phase of binary neutron star mergers may also set constraints on $M_{\rm max}$ \cite{2020arXiv200500482C}.

An extension of the empirical relation Eq. (\ref{eq:mthr}) as  a more accurate, bilinear fit of the form \begin{equation}
M_{\max }\left(M_{\text {thres }}, \widetilde\Lambda_{\text {thres }}\right)=0.632 M_{\text {thres }}-0.002 \widetilde\Lambda_{\text {thres }}+0.802,
\end{equation}
 where \(\widetilde\Lambda_{\text {thres }}=\widetilde\Lambda\left(M_{\text {thres }} / 2\right)\), was recently presented in\cite{2020arXiv200400846B}, and a unique signature of strong phase transitions was found in the \(\widetilde\Lambda_{\text {thres }}\) vs. \(M_{\text {thres
}}\) parameter space.

\subsection{Future constraints from post-merger NS oscillations}
\label{post}

\begin{figure*}
\begin{center}
\resizebox{0.7\textwidth}{!}{
  \includegraphics{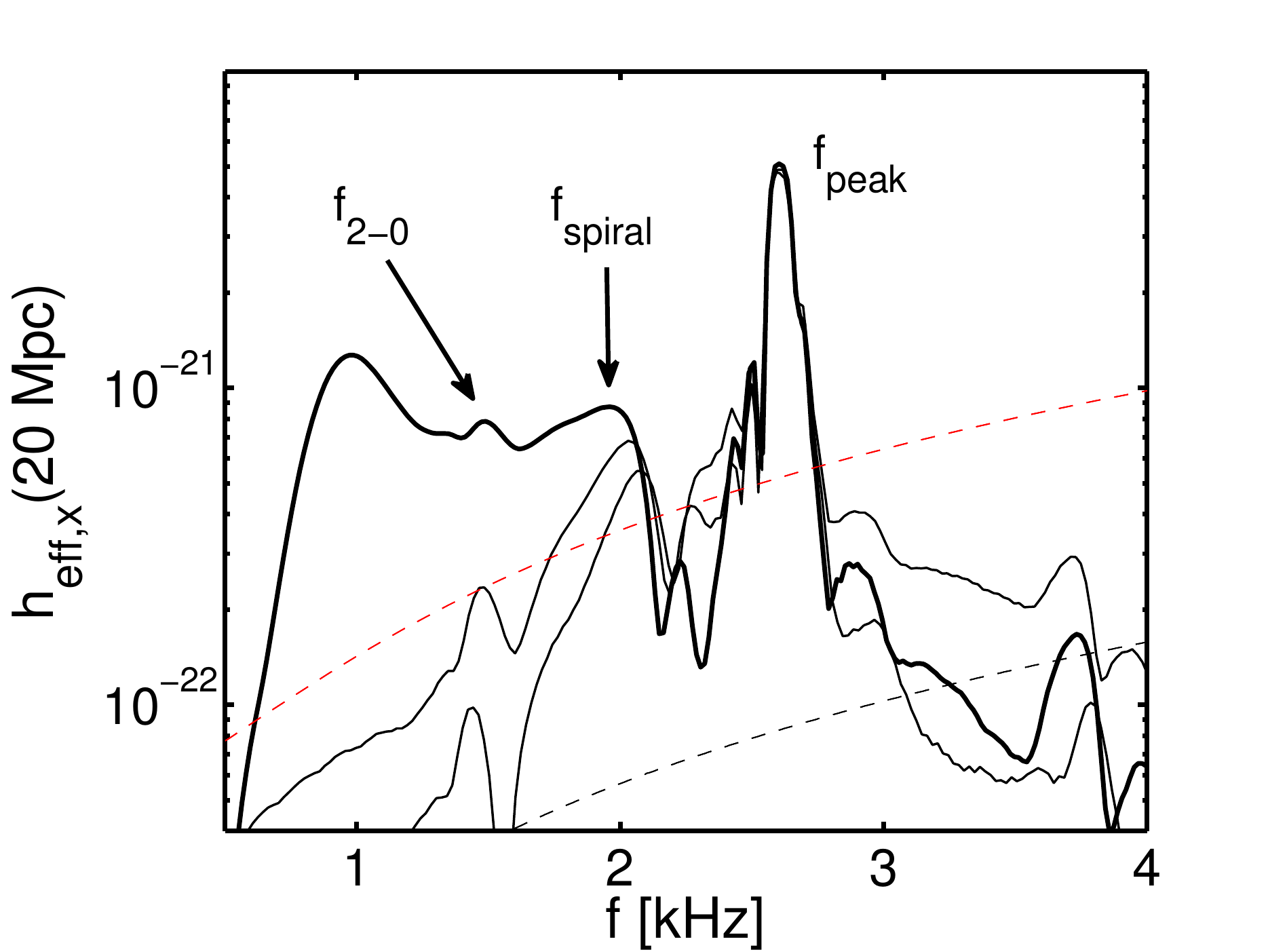}
}
\end{center}
  \caption{GW spectrum of a 1.35-1.35~$M_\odot$
merger described by the DD2 EoS \cite{Hempel2010,Typel2010} viewed along the polar
direction at a distance of 20~Mpc.  The frequencies $f_\mathrm{peak}$, $f_\mathrm{spiral}$
and $f_{2-0}$ are distinct features of the postmerger phase and can be
associated with particular dynamical effects in the remnant. The thin solid
lines display the GW spectra when the inspiral phase is ignored (with different
cutoffs), revealing that the peaks are generated in the postmerger phase.
Dashed lines show the expected design sensitivity curves of Advanced LIGO
\cite{Harry2010} (red) and of the Einstein Telescope \cite{Hild2010} (black).
Notice that the $f_{2+0}$
frequency is also present (at about 3.7\,kHz), but it is very weak to be
detectable even with the Einstein Telescope. Figure  from \cite{Bauswein2016}.}
  \label{fig:spectrum}
\end{figure*}

A likely outcome of a NS merger is the formation of a metastable,
differentially rotating NS remnant. The expected GW spectrum of such a case
(in terms of the
effective GW amplitude $h_{\rm eff}=\widetilde{h}(f)\cdot f$, where $f$ is frequency
and $\tilde h(f)$ the Fourier transform), is shown in Fig.~\ref{fig:spectrum} for a NS merger simulation of two stars with a mass
of 1.35~$M_\odot$ each, assuming the EOS DD2. \cite{Hempel2010,Typel2010} (notice that these hydrodynamical
simulations start only a few orbits before merging).  
The postmerger spectrum exhibits several distinct peaks in the kHz range~\cite{Stergioulas2011,Hotokezaka2013a,Takami2014,Kastaun2015,Bauswein2015,Palenzuela2015,Takami2015,Bauswein2016,DePietri2016,Foucart2016,Rezzolla2016,Clark2016,Dietrich2017a,2017CQGra..34c4001F,Maione2017},
which originate from certain oscillation modes and other dynamical processes
in
the postmerger remnant.  There is a dominant oscillation
frequency $f_\mathrm{peak}$ (also denoted as $f_2$), which  typically  has
the highest signal-to-noise ratio (SNR) of all distinct postmerger features. In addition, there are
secondary peaks, denoted as $f_\mathrm{spiral}$
and $f_{2-0}$, that are above the aLIGO/aVIRGO noise level  and additional
peaks at higher frequencies, which are less likely to be observed, even with
third-generation detectors. Understanding the physical mechanisms generating
these different features
is essential for the detection and interpretation of postmerger GW signals.

 An effective method for analyzing oscillation modes of rotating stars, based
on a Fourier extraction of their eigenfunctions from simulation data, was
presented in \cite{Stergioulas2004} and applied to NS merger remnants in \cite{Stergioulas2011}. 
Throughout the star, the Fourier spectrum exhibits a discrete dominant frequency $f_\mathrm{peak}$, with an extracted eigenfunction that has a clear $m=2$ quadrupole structure and no nodal lines in the 
radial direction. Thus,  $f_{\rm peak}$ is produced by the fundamental quadrupolar fluid oscillation mode 
of the post-merger remnant, a result confirmed by hydrodynamical simulations of the late-time remnant 
 with the $m=2$ fundamental mode  added as as perturbation\cite{Bauswein2016}.

\begin{figure*}
\begin{center}
\resizebox{0.8\textwidth}{!}{
  \includegraphics{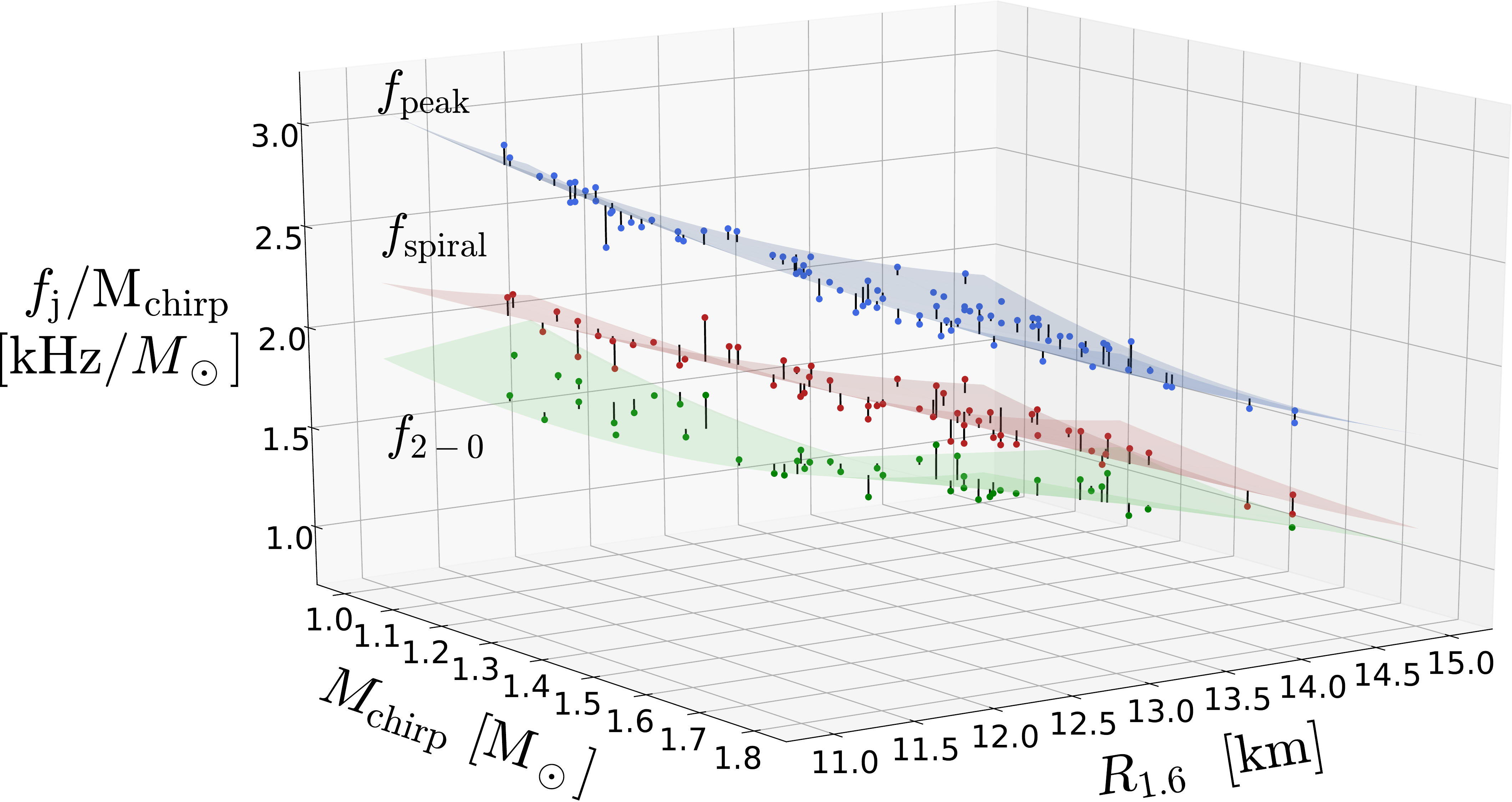}
}
\end{center}
  \caption{Empirical surfaces for the three main post-merger frequencies, as a function of $M_{\rm chirp}$ and $R_{1.6}$. The surfaces are shown only in regions where data points are available. Figure
from Ref\cite{Vretinaris2020}.}
  \label{fRMsurfaces2gether}
\end{figure*}
 
The collision of the two stars also excites the   fundamental
quasi-radial mode of oscillation in the remnant, whose frequency 
we denote by  $f_0$. Since these oscillations are nearly spherically symmetric even in a rotating remnant, the quasi-radial
mode produces only weak GW emission (typically at a frequency where the spectrum
is still dominated by the inspiral phase). However, both the quadrupole and the quasi-radial oscillations
have a large initial amplitude, and thus there exist non-linear couplings
between them. At second order in the perturbation, the coupling of the two modes results in 
the appearance of\textit{ daughter modes}, here with 
{\it quasi-linear combination frequencies}\cite{Stergioulas2011} $f_{2\pm0}=f_\mathrm{peak}\pm f_0$.

At frequencies below $f_\mathrm{peak}$, there is at least one more pronounced secondary peak, 
which (when present) typically appears in between $f_\mathrm{peak}-
f_0$ and  $f_\mathrm{peak}$, as shown in Fig.~\ref{fig:spectrum}.
This secondary peak, denoted by $f_{\rm spiral}$, is generated by the orbital motion of two 
bulges that form right after merging  \cite{Bauswein2015}.
 This feature is strongest in equal-mass binaries, appearing as two small spiral arms. Matter in these bulges or spiral arms
cannot follow the faster rotation of the quadrupole pattern of the inner core; instead, the antipodal bulges orbit 
the central remnant with a smaller orbital frequency.
The structure is transient and dissolves within a few milliseconds.
Notice that the $f_{2+0}$
frequency can also be present in the spectrum, but it is typically too weak to be
detectable even with the Einstein Telescope (although this could become possible with dedicated high-frequency detectors).%

\begin{figure}
\begin{center}
\resizebox{0.7\textwidth}{!}{
  \includegraphics{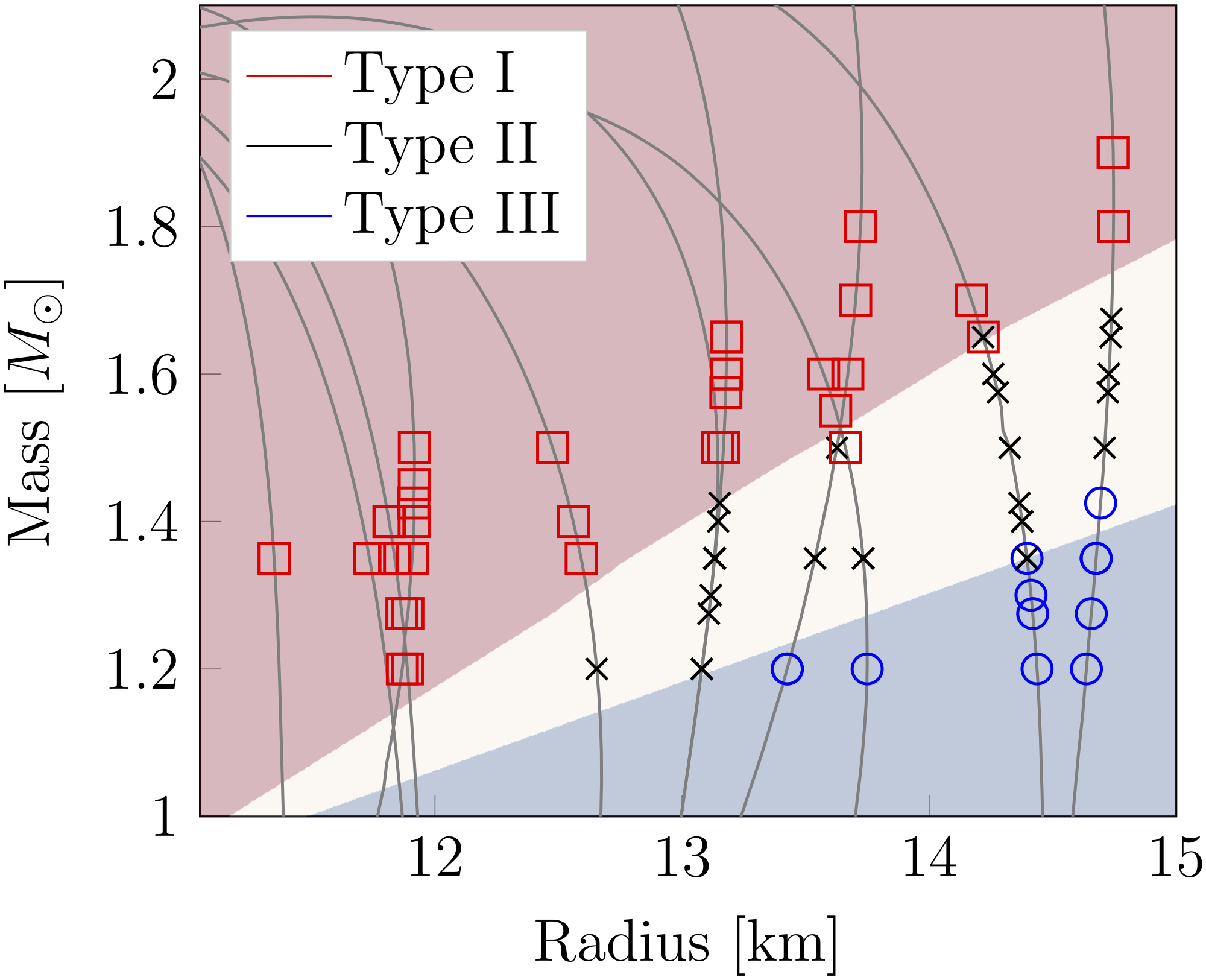}
}
\end{center}
\caption{Spectral classification of the postmerger GW emission, as obtained
by a machine-learning algorithm, verifying the classification introduced
in\cite{Bauswein2015}. The
classification is shown in the mass vs. radius parameter space of nonrotating
neutron star models, constructed with various EOSs and masses.
A clustering algorithm separates the models into three different types (shown
as red squares for Type I, black $\times$ for Type II\ and blue circles for
Type III). Then, a supervised-learning classification algorithm locates the
borders between the three different types in this parameter space. Figure
from Ref\cite{Vretinaris2020}.}
\label{fig:classification}
\end{figure}

In the range of total masses $2.4 M_\odot \leq M_{\rm tot} \leq 3.0 M_\odot$,
the secondary peaks appear in distinct frequency ranges\cite{Bauswein2015}:
$f_{\rm peak} -
1.3 {\rm kHz} \leq f_{2-0} \leq f_{\rm peak} - 0.9 {\rm kHz} $, while $f_{\rm
peak} - 0.9 {\rm kHz} \leq f_{\rm spiral} \leq f_{\rm peak} - 0.5{\rm kHz}
$. All three distinct post-merger GW frequencies can be described by empirical
relations with small scatter.  Recently, multivariate  empirical relations
were constructed, where each frequency  is given as a function of the chirp
mass $M_{\rm chirp}$ and the radius of a $1.6\,M_\odot$ nonrotating star\cite{Vretinaris2020},
see Fig. \ref{fRMsurfaces2gether}.

Examining the post-merger GW spectrum for equal-mass (or nearly equal-mass)
binaries, one can arrive at the following \textit{spectral classification}\cite{Bauswein2015}:
 
\begin{itemize}
\item {\bf Type I}: The $f_{2-0}$ peak is the strongest secondary feature,
while the $f_\mathrm{spiral}$ peak is suppressed or hardly visible. This
behavior is found
for mergers with relatively \textit{high total binary masses} and \textit{soft
EOSs}. 
\item {\bf Type II}: Both secondary features $f_{2-0}$
and $f_\mathrm{spiral}$ are clearly present and have roughly comparable strength.

\item {\bf Type III}: The $f_\mathrm{spiral}$ peak is the strongest secondary
feature, while the   $f_{2-0}$ peak is either strongly suppressed or even
absent.
This
behavior is found
for mergers with relatively \textit{low total binary masses} and \textit{stiff
EOSs.}
\end{itemize}
This classification scheme was reproduced in Ref. \cite{Vretinaris2020}
using machine-learning algorithms, see Fig. \ref{fig:classification}. For
a given EOS, different spectral types may occur, depending on the
total mass of the binary.
For asymmetric mass ratios of $q\sim 0.7$ the above classification
scheme has to be modified (for example, in such asymmetric
cases the $f_{\rm spiral}$ secondary peak will be considerably weaker).

In~\cite{Abbott2019} an unmodelled data analysis
search was performed to extract the postmerger GW emission. No signal
was found, as is expected for the given distance of the event and the
sensitivity of the instruments during the observations.
Improving the sensitivity of detectors by a factor of a few, however, 
may allow the detection of postmerger
gravitational-wave emission at a distance of few tens of Mpc~\cite{Clark2014,Clark2016,Chatziioannou2017,Torres-Rivas2019}.  With
this sensitivity, the masses of the individual
stars could be measured with a precision
of a percent at the same distances~\cite{Rodriguez2014,Farr2016,Abbott2017},
if the accuracy of mass measurements scales
roughly with $(\mathrm{SNR})^{-1}$ or $(\mathrm{SNR})^{-1/2}$. For binaries at larger distances, stacking 
a large number of individual signals may allow the detection of the post-merger
phase with third-generation detectors.\cite{Bose2018,Yang2018}

\begin{figure}
\begin{center}
\resizebox{0.7\textwidth}{!}{
  \includegraphics{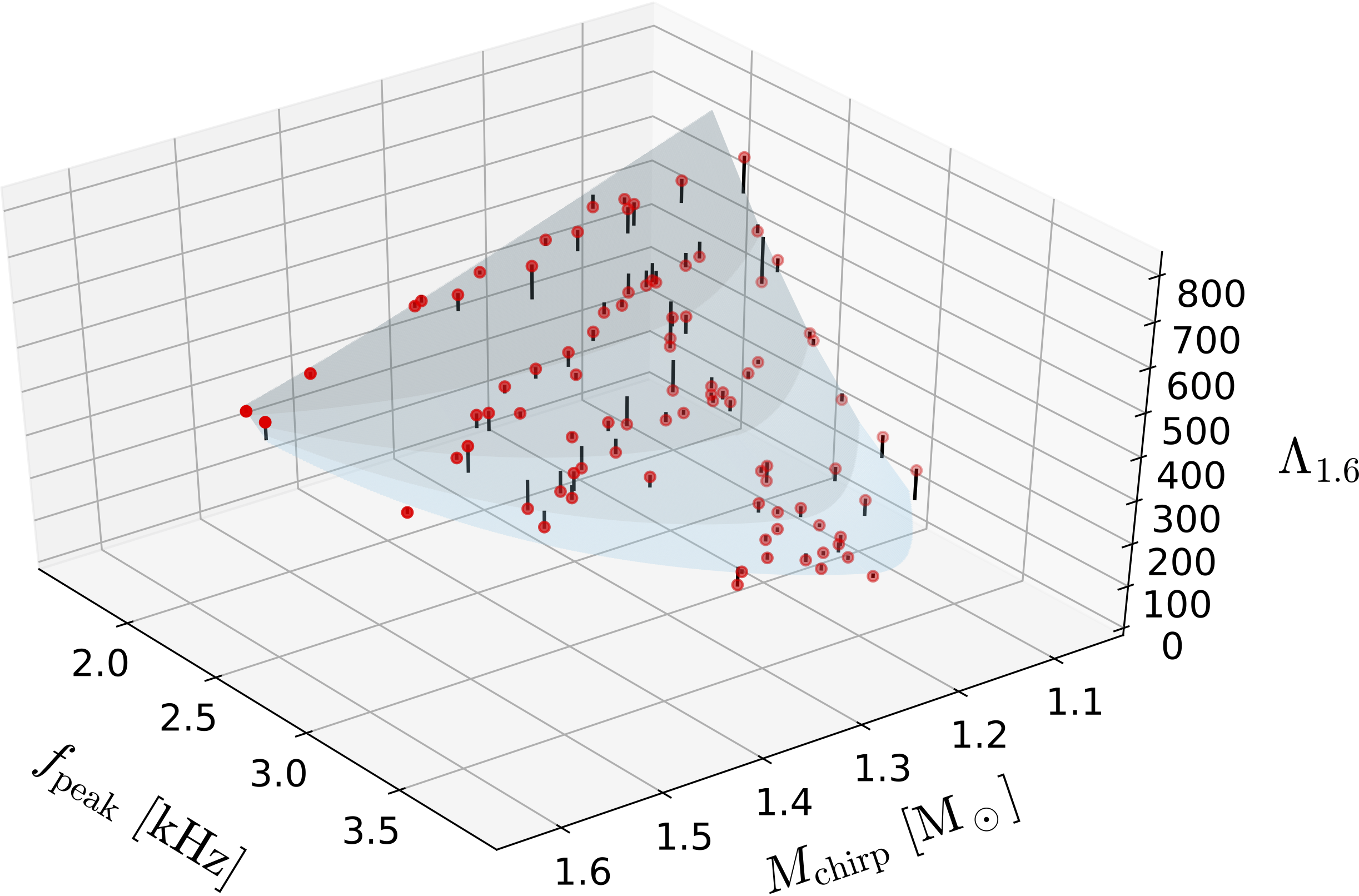}
}
\end{center}
\caption{ Multivariate empirical relation for $\Lambda_{1.6}$ as a function
of $f_{\rm peak}$ and $M_{\rm chirp}$. Figure
from Ref\cite{Vretinaris2020}.}
 \label{fig:LfM:Ltilde2D}
\end{figure}

The peak frequency
$f_\mathrm{peak}$ of 1.35-1.35~$M_\odot$ mergers shows a clear correlation
with the radius $R_{1.35}$ of a nonrotating NS with 1.35~$M_\odot$ (see Fig.~4
in \cite{Bauswein2012} and Fig.~12 in \cite{Bauswein2012a}). Similar tight
correlations exist for other fiducial masses  (see Figs.~9 to 12 in \cite{Bauswein2012a}).
The tightest relation for a 1.35-1.35~$M_\odot$ is with the radius $R_{1.6}$. 
With $f_{\mathrm{peak}}$ in kHz and $R_{1.6}$ in km, the relation has the form
\begin{equation}
f _ { \mathrm { peak } } = \left\{ \begin{array} { l l } 
           { - 0.2823 \cdot R _ { 1.6 } + 6.284, } & { \textrm { for } f
_ { \rm peak  } < 2.8 \mathrm { kHz } }, \\ 
           { - 0.4667 \cdot R _ { 1.6 } + 8.713, } & { \textrm { for } f
_ { \rm peak  } > 2.8 \mathrm { kHz } }. \end{array} \right.
\end{equation}
For $R_{1.6}$ the maximum scatter of this relation is less than 200~m. 

For other fixed binary masses, e.g.\ 1.2-1.2~$M_\odot$, 1.2-1.5~$M_\odot$
or 1.5-1.5~$M_\odot$ mergers, similar scalings between $f_\mathrm{peak}$
and NS radii exist~\cite{Bauswein2012a} and a single relation, scaled by
the total mass, is \cite{Bauswein2016}
\begin{equation}
f _ { \rm peak  } / M _ { \mathrm { tot } } = 0.0157 \cdot R _ { 1.6 } ^
{ 2 } - 0.5495 \cdot R _ { 1.6 } + 5.5030,
\end{equation}
(see~\cite{Bernuzzi2015} for a similar rescaling but with the tidal coupling constant).

A multivariate extension\cite{Vretinaris2020} of the above empirical relations yields the radius of nonrotating neutron stars at a specified mass as a function of two observable variables, $R_{\rm x}=R_{\rm x}(f_{\rm peak}, M_{\rm chirp})$,  
where $\rm x$ stands for the value of the mass in solar masses ($f_{\rm peak}$ obtained from the post-merger GW spectrum and $M_{\rm chirp}$ obtained from the inspiral  phase). As an example, for the case of $M = 1.6 M_\odot$ 
the empirical relation for the radius is
\begin{equation}
  \begin{split}
    R_{1.6}= 35.442 -13.46 M_{\mathrm{chirp}} -9.262 f_{\mathrm{peak}}/M_{\mathrm{chirp}}\\
+3.118 M_{\mathrm{chirp}}^2 +2.307 f_{\mathrm{peak}} +0.758 \left( f_{\mathrm{peak}}/M_{\mathrm{chirp}}
\right)^2,
  \end{split}
  \label{Combined_RfM_R16_e0}
\end{equation}
with a maximum residual of 0.654 km and $R^2=0.954$.  This relation was constructed using an extended set of simulations and available GW catalogues, that included equal and unequal mass binaries.

Multivariate empirical relations can also be constructed for the dimensionless tidal deformability\cite{Vretinaris2020}. For example, for $M = 1.6 M_\odot$  one obtains
\begin{equation}
   \Lambda_{1.6} = 2417 + 770.2 M_{\rm chirp} - 1841 f_{\rm peak}
+ 262.9 f_{\rm peak}^2,
\label{L1.6}
\end{equation}
with maximum residual of 99.85 and $R^2=0.964$, see Fig. \ref{fig:LfM:Ltilde2D}.

Additional empirical relations for setting EOS constraints have been presented e.g. in Refs.\cite{Hotokezaka2013a,Bernuzzi2015,Rezzolla2016,2019PhRvD.100d4047T,2019PhRvD.100j4029B}. A one-armed instability was studied in Refs.\cite{2015PhRvD..92l1502P,2016PhRvD..93b4011E,2016CQGra..33x4004E,2016PhRvD..94f4011R,2019PhRvD.100l4042E} and the excitation of inertial modes (under the assumption of low viscosity in the remnant) was studied in Refs.\cite{2018PhRvL.120v1101D,2020PhRvD.101f4052D}  The detection of phase transitions through GW\ observations of the post-merger phase was considered in\cite{2017ApJ...842L..10R,2019PhRvL.122f1102B,2019ApJ...881..122D,2019Univ....5..156H,2019PhRvL.122f1101M,2019arXiv191209340W,2020EPJA...56...59M}. For the impact of magnetic fields on the post-merger phase, see the review\cite{2020arXiv200307572C} and references therein and for the effect of a possible strong turbulent viscosity, see the review\cite{sh19} and references therein, as well as the recent study using a calibrated subgrid-scale turbulence model\cite{2020arXiv200509002R}. Because of the high scientific return expected from observing the post-merger phase with gravitational waves, designs for new GW detectors, which will be dedicated to operate with enhanced sensitivity in the kHz regime, have been presented.\cite{2019PhRvD..99j2004M,2019arXiv191206305B,2020arXiv200202637A}
For a large number of sources at cosmological  distances, the cumulative information on radius will be dominated by sources at  $z\sim 1$.\cite{2020arXiv200411334H}

\section{Measuring Hubble's Constant}
\label{s:hubble}
 
Hubble's constant $H$ is the fractional rate at which the universe expands.  
For nearby galaxies whose relative motion tracks the expansion, 
the present value of $H$ is given by
\be
  H_0 = v/d, 
\label{e:H}\ee
where $d$ is the proper distance between the galaxies and $v$ their present relative velocity, 
the present rate of change of their proper distance. 
The universe appears to be nearly spatially flat, and for a spatially flat geometry, 
Eq.~\eqref{e:H} is exact for galaxies that track the Hubble flow. 
  
Late universe measurements of $H_0$ conflict with early-universe measurements from the cosmic microwave background at $z\approx 1100$. The early-universe measurements by the WMAP \cite{wmap13} and Planck satellites \cite{Planck2020} of the cosmic microwave background give
\be
  H_0 = 69.32\pm 0.80\ \rm km/s/Mpc \mbox{ and } H_0 = 67.66\pm 0.42 \rm km/s/Mpc, 
\ee
respectively. Here $H_0$ is not directly measured; its value relies on the $\Lambda$CDM model, whose six parameters fit with high precision the cosmological data.

In contrast, on the astrophysical side, the SH$_0$ES collaboration \cite{riess19} uses the brightness of type Ia supernovae (SNe Ia) as standard candles to find the larger value 
\[
  H_0 = 74.03\pm 1.42\ \rm km/s/Mpc.
\]
This measurement rests on the Leavitt relation between period and luminosity of Cepheid variables, now found directly by parallax \cite{riess18a,riess18b} and by geometrical determination of distance to the Large Magellanic Cloud (see \cite{pietrzynski19} for a current determination and for other references).  Independent ways to determine $H_0$ include water maser lines in imaged thin disks about galactic black holes, which give a value \cite{megamaser20} 73.9$\pm$ 3.0 km/s/Mpc; and measurements of time delays of different images from the same from multiply-imaged quasar giving \cite{wong20} $73.3_{-1.8}^{+1.7}$. The {\sl Hubble tension}, the discrepancy between the early- and late-universe values, is now above 4$\sigma$.\cite{riess19} \footnote{However, a recent late-universe measurement by the Carnegie-Chicago Hubble Project calibrates Type Ia Sne by the brightness and luminosity of the tip of the red-giant sequence, finding $H_0 = 69.8 \pm 0.8({\rm statistical})\pm 1.7({\rm systematic})$ km/s/Mpc. }   

The key question is whether the discrepancy indicates new physics beyond the $\Lambda$CDM model (see, for example, Knox and Millea \cite{knox20}) or is due to unknown systematic errors.  
In 1986, Schutz \cite{schutz86} pointed out that the binary-inspiral waveform
could be used to give an independent measure of $H_0$ if one could identify a host galaxy 
with measured redshift (see also Holz and Hughes\cite{hh05}, Nissanke \etal\cite{nissanke13} 
and references therein).  Remarkably, the first observed binary neutron star inspiral 
gives exactly that measurement.  

The relation is derived as follows.
The distance $D$ to the binary is given in terms of three waveform observables, 
the frequency $\omega$, its first time derivative, $\dot\omega$ and its amplitude $h$.  
At lowest post-Newtonian order, three equations relate these quantities to $D$: 
\begin{enumerate} 
\item The radiation is quadrupole, and the frequency of quadrupole radiation
is related to the frequency $\Omega$ of the orbit by 
\be
 \omega = 2\Omega.
\ee
\item For a circular orbit, the quadrupole formula \eqref{e:amplitude} for the rms amplitude 
of the waveform can be written 
as 
\be
  \bar h \propto \frac ED.
\label{e:hED}\ee
where $E$ is the Newtonian energy of the binary.
\item
Finally, the rate $\dot E$ at which the system loses energy to gravitational waves 
is the energy flux \eqref{e:Edotq2} at the distance $D$,  
\be 
   \dot E \propto \langle\dot h^2\rangle D^2 = \omega^2 \bar h^2 D^2. 
\label{e:Edot}\ee
\end{enumerate}
In each case, the constant of proportionality depends only on $G$ and $c$: It is 
purely numerical in gravitational units. Then, with $k$ again a constant of this form,
\be
  \frac{\dot\omega}\omega = \frac32 \frac{\dot E}E = \frac1k \frac{\omega^2 \bar h^2 D^2}{\bar h D},  
\ee
where Eqs.~\eqref{e:hED} and \eqref{e:Edot} were used in the last equality 
to write $E$ and $\dot E$, respectively, in terms of $h$ and $D$.  Thus 
\be
   D = k \frac{\dot\omega}{\omega^3 \bar h}. 
\ee
The simplicity of this relation comes from that fact that $\dot E/E$ 
is independent of the two masses.   With the constants included and $\omega=2\pi f$, 
the relation has the form 
\be
   D = 780 \frac{\dot f_{100}}{f_{100}^3 \bar h_{23}}\ \rm Mpc, 
\ee
where $f_{100} = f/(100\rm Hz)$, $\bar h_{23} = 10^{23}\bar h$, and $\dot f/f$ is given in 
s$^{-1}$.  

Note that, in this relation, the distance $D$ of Eqs.~\eqref{e:hED} and \eqref{e:Edot} 
is the {\it luminosity distance} to the 
binary, related to the energy flux $\tt f$ of gravitational waves (or of light) by 
$L\equiv \dot E = 4\pi D^2\tt f$.  For a spatially flat geometry, 
the present proper distance $d$ of Hubble's law is given by
\be
  d = D/(1+z),
\ee
with $z$ the redshift.      

For $GW170817$, the distance $D$, when found solely from the gravitational-wave observation, 
has the value $D = 43.8^{+2.9}_{-6.9}\ \rm Mpc$ \cite{hubble_gw17}.  The redshift, adjusted to account 
for the galaxy's peculiar velocity (its deviation from the Hubble flow), is 
$z = (1.01\pm 0.06)\times 10^{-2}$, corresponding to a Hubble-flow velocity 
$v=3.02\pm 17$ km/s, and giving  
\be
  H_0 = 70^{+12}_{-8}\rm km/s/Mpc. 
\ee
The central value is consistent with the current electromagnetic measurements, 
but we need to observe of order 40 systems with this accuracy to reduce the statistical error to 
a level that could discriminate between the values from the cosmic background and 
from local electromagnetic distance determinations.

  Messenger and Read\cite{mr12}  noticed that, in NS-NS inspiral, the tidal terms 
add to the waveform an additional measurable parameter. This, in principle, allows 
one to measure both luminosity distance and redshift, thereby determining the 
Hubble constant from the waveform alone. Although well beyond the scope of the 
current detectors, it may be within reach of the next generation.

\section*{Acknowledgments}

We are grateful to A. Bauswein for comments on the manuscript and to the referee for many corrections. N.S. is supported by the ARIS facility of GRNET in Athens (SIMGRAV, SIMDIFF\ and BNSMERGE allocations) and the Aristoteles Cluster at AUTh, as well as by the COST actions CA16214 (PHAROS), CA16104 (GWVerse), CA17137 (G2Net) and CA18108 (QG-MM).

\bibliographystyle{ws-ijmpd}
\bibliography{b20}

\end{document}